\documentclass[nohyper,11pt,letterpaper]{JHEP3}
\pdfoutput=1
\usepackage{epsf,amsfonts,amssymb,epsfig,bm,amsmath,yfonts}
\usepackage{graphicx}
\usepackage{bm}
\usepackage{amsmath}

\newcommand{\be}{\begin{equation}}
\newcommand{\ee}{\end{equation}}
\newcommand{\bea}{\begin{eqnarray}}
\newcommand{\eea}{\end{eqnarray}}
\newcommand{\ba}{\begin{eqnarray}}
\newcommand{\ea}{\end{eqnarray}}
\newcommand{\nn}{\nonumber \\}
\newcommand{\eqn}[1]{(\ref{#1})}
\newcommand{\beq}{\begin{equation}}
\newcommand{\eeq}{\end{equation}}
\newcommand{\beqa}{\begin{eqnarray}}
\newcommand{\eeqa}{\end{eqnarray}}
\newcommand{\beqar}{\begin{eqnarray*}}
\newcommand{\eeqar}{\end{eqnarray*}}

\newcommand{\equ}[1]{(\ref{#1})}

\newcommand{\sac}{\, , \qquad}

\def\nc {N_\mt{c}}
\def\nf {N_\mt{f}}

\def\t6 {T_\mt{D6}}

\def\gym {g_\mt{YM}}

\newcommand{\mq}{M_\mt{q}}      

\newcommand{\mt}[1]{\textrm{\tiny #1}}

\newcommand{\tildef}{\tilde{f}}

\newcommand{\tdec}{T_\mt{c}}
\newcommand{\mmes}{M_\mt{mes}}

\newcommand{\prt}{\partial}

\newcommand{\vlim}{v_\mt{lim}}

\newcommand{\calj}{{\cal J}}
\newcommand{\calf}{{\cal F}}
\newcommand{\caln}{{\cal N}}
\newcommand{\cala}{{\cal A}}

\newcommand{\eps}{\epsilon}

\title{Cherenkov mesons as in-medium quark energy loss}

\author{Jorge Casalderrey-Solana,$^1$ Daniel Fern\'andez$^2$ 
and David Mateos$^{3,2}$ \\
$^1$Physics Department, Theory Unit, CERN, CH-1211 Gen\`eve 23, Switzerland \\ \\
$^2$Departament de F\'\i sica Fonamental \&  Institut de Ci\`encies del Cosmos (ICC), Universitat de Barcelona, Mart\'{\i}  i Franqu\`es 1, E-08028 Barcelona, Spain \\ \\
$^3$Instituci\'o Catalana de Recerca i Estudis Avan\c cats (ICREA), Passeig Llu\'\i s Companys 23, E-08010, Barcelona, Spain}

\abstract{We recently showed that a heavy quark moving sufficiently fast through a quark-gluon plasma may lose energy by Cherenkov-radiating mesons \cite{prl}. Here we review our previous holographic calculation of the energy loss in ${\cal N}=4$ Super Yang-Mills and extend it to longitudinal vector mesons and scalar mesons. We also discuss phenomenological implications for heavy-ion collision experiments. Although the Cherenkov energy loss is an ${\cal O}(1/\nc)$ effect, a ballpark estimate yields a value of $dE/dx$ for $\nc=3$ which is comparable to that of other mechanisms.}


\preprint{CERN-PH-TH/2010-214 \\ ICCUB-10-055}

\begin{document}

\section{Introduction}
With the advent of the Large Hadron Collider (LHC) the field of heavy-ion collisions (HIC) enters a new era. The center-of-mass energy per nucleon in LHC collisions, $\sqrt{s_\mt{NN}} \simeq 5.5$ TeV, is almost 30 times larger than that of the most energetic collisions at the Relativistic Heavy Ion Collider (RHIC). The highest temperature of the quark-gluon plasma (QGP) created in RHIC experiments is approximately $T_\mt{RHIC} \simeq 2 \tdec$, with $\tdec \simeq 175$ MeV the deconfinement temperature of Quantum Chromodynamics (QCD). Despite the large increase in the collision energy, this is expected to lead only to a moderate increase in the plasma temperature at the LHC \cite{lhc}, i.e. $T_\mt{LHC} \simeq (3-4) \tdec$.\footnote{A rough estimate is obtained by assuming that the temperature scales as the fourth root of the energy density.} In contrast, 
high-energy partons originating from hard initial collisions will be copiously produced at the LHC. This will allow the study of quarks and gluons in the 100 GeV range, an order of magnitude larger than that at the RHIC.
 
Experimentally, extremely valuable information is obtained by analyzing the energy loss of these energetic partons as they travel through the QGP. In order to use this information to learn about the plasma, a theoretical, quantitative understanding of the different mechanisms of parton energy loss is needed. Several such mechanisms have been previously studied, both in QCD itself \cite{radiative} and in the context of the gauge/gravity duality \cite{strong}.

A remarkable conclusion from the RHIC experiments \cite{rhic} is that the QGP does not behave as a weakly coupled gas of quarks and gluons, but rather as a strongly coupled fluid \cite{fluid}. Because of the moderate increase in the temperature and the logarithmic running of the QCD coupling constant, a qualitatively rather similar behaviour may be expected for the QGP at the LHC. This makes it particularly important to understand mechanisms of parton energy loss that may operate at strong coupling. We recently uncovered one such mechanism \cite{prl} whereby a sufficiently fast heavy quark traversing a strongly coupled plasma loses energy by Cherenkov-radiating in-medium mesons. 

The analysis in \cite{prl} showed that this mechanism takes place in all strongly coupled, large-$\nc$ gauge theory plasmas with a gravity dual. The argument is so simple that we reproduce it in section \ref{universal} for completeness. This section emphasizes the universality of the mechanism, since no reference to a specific model is necessary.

Ref.~\cite{prl} also performed a quantitative analysis in the simple example of a quark moving through the ${\cal N} =4$ super Yang-Mills (SYM) plasma. The quark Cherenkov-radiates both vector and scalar mesons. The rate of energy loss into the transverse modes of the vector mesons was calculated in \cite{prl}, and again we reproduce it here for completeness. The vector mesons in question are massive, and thus they also possess a longitudinal mode. Here we extend the calculation of \cite{prl} and obtain the rate of energy loss into longitudinal vector mesons and scalar mesons. The result for the former is qualitatively similar to that for the transverse modes, whereas the result for scalar mesons displays some qualitative differences. 

Ref.~\cite{prl} presented a rather preliminary exploration of the potential implications of these results for HIC experiments. Here we elaborate on that discussion and extend it to include possible implications of the new results presented in this paper.

\section{A universal mechanism of quark energy loss}
\label{universal}
The reason that the mechanism we are going to describe is universal is that it only relies on two universal features of the gauge/gravity duality:\footnote{In the limit 
$\nc, \gym^2 \nc \rightarrow \infty$.} (i) the fact that the deconfined phase of the gauge theory is described by a black hole geometry on the gravity side \cite{Witten}, and (ii) the fact that a finite number $\nf$ of quark flavours is described by $\nf$ D-brane probes \cite{Karch-Randall} -- see fig.~\ref{cherenkov-with-quark}. 
\FIGURE{
\includegraphics[scale=.70]{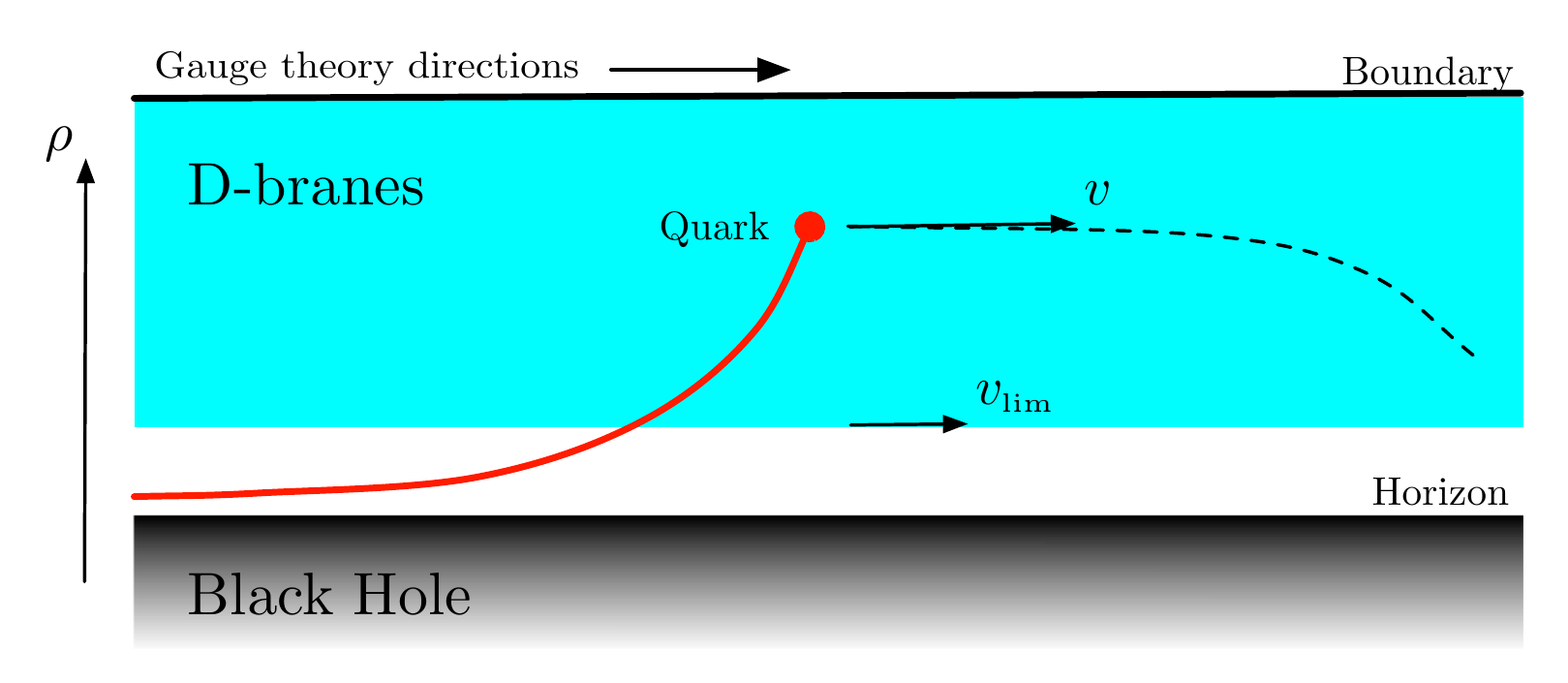}
\caption{D-branes and an open string in a black hole geometry.} 
\label{cherenkov-with-quark}}

In addition to the gauge theory directions, the gravity description always includes a radial direction (denoted by $\rho$ in fig.~\ref{cherenkov-with-quark}) which is dual to the gauge theory energy scale. The radial position of the horizon is proportional to the plasma temperature $T$. The D-branes extend in the radial direction down to a minimum value proportional to the quark mass $\mq$. 

As it is intuitively clear, for $\mq$ sufficiently larger than $T$ the D-branes sit completely outside the horizon \cite{MMT, MMT2, dpdq, towards, dpdqdqbar}.\footnote{In contrast, as the ratio $\mq/T$ decreases, a first-order phase transition eventually occurs and a part of the branes falls through the horizon. See section \ref{quarks}.} In this phase, scalar and vector gauge theory mesons are described by small, normalizable fluctuations of scalar and vector fields propagating on the branes,
whose low-energy dynamics is governed by a Maxwell-like theory. The spectrum of these fluctuations is discrete and gapped, which means that stable heavy meson states exist in the plasma. In other words, sufficiently heavy mesons survive deconfinement, in agreement with lattice and potential model predictions for QCD \cite{quarkonium}.

Consider now the in-medium dispersion relation $\omega(q)$ for these heavy mesons, 
where $\omega$ and $q$ are the energy and the spatial three-momentum of the meson, respectively. As an illustrative example, the dispersion relations for vector and scalar mesons in the ${\cal N}=4$ SYM plasma\footnote{With quarks introduced as D7-branes; see below for details.} are depicted in fig.~\ref{dispersion}. 
\FIGURE{
\begin{tabular}{cc}
\,\,\,\,\,\,\,\,\,\,\,\,
\includegraphics[width=0.43 \textwidth]{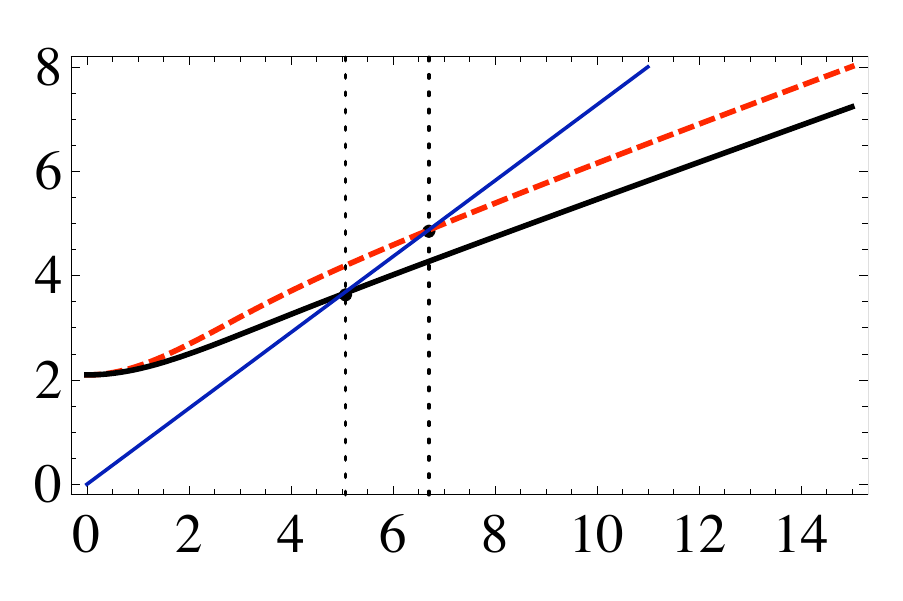}  
\put(-105,-4){$q/\pi T$}
\put(-200,50){\rotatebox{90}{{$\omega/\pi T$}}}
&
\,\,\,\,\,\,\,\,\,\,\,\,
\includegraphics[width=0.43 \textwidth]{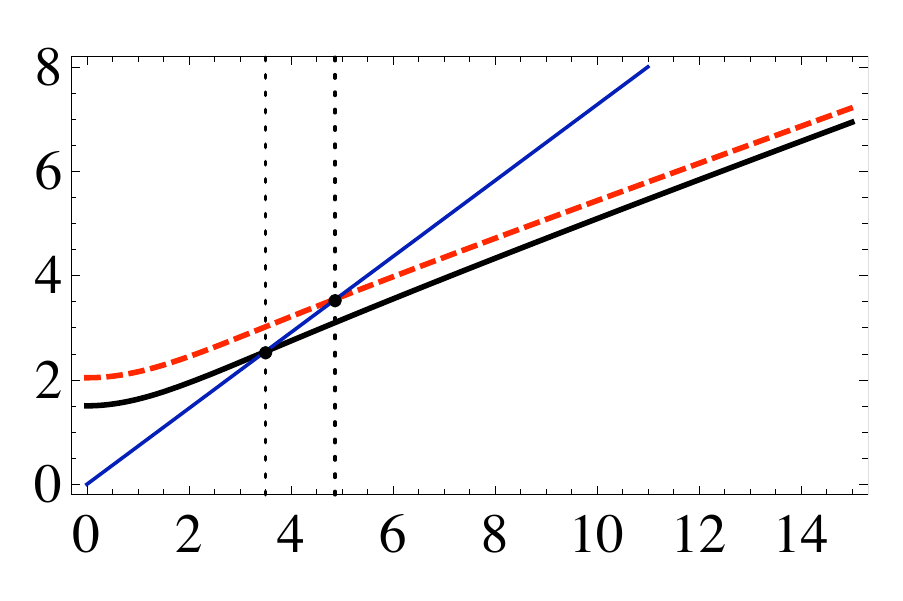} 
\put(-105,-4){$q/\pi T$}
\put(-200,50){\rotatebox{90}{{$\omega/\pi T$}}}
\end{tabular}
\caption{Left: Dispersion relation for the transverse (black, continuous curve) and longitudinal (red, dashed curve) $n=0$ modes of a heavy vector meson with $\vlim = 0.35$ in the ${\cal N}=4$ SYM plasma. Right: Analogous curves for a scalar (black, continuous curve) and pseudoscalar (red, dashed curve) meson. In both plots the blue, continuous straight lines correspond to $\omega = v q$ with $\vlim<v<1$. The black, dotted, vertical lines mark the crossing points between the meson dispersion relations and the blue lines.} 
\label{dispersion}}

As $q \rightarrow \infty$, the DR becomes linear: $\omega(q) \sim \vlim q$, with $\vlim < 1$. This subluminal limiting velocity, which is the same for all mesons, can be easily understood in the gravitational description \cite{MMT2}. Since highly energetic mesons are strongly attracted by the gravitational pull of the black hole, their wave-function is very concentrated at the bottom of the branes. Consequently, their velocity is limited by the local speed of light at that point, $\vlim$ (see fig.~\ref{cherenkov-with-quark}). Because of the black hole redshift, $\vlim$ is lower than the speed of light at infinity (i.e. at the boundary), which is normalised to unity. In the gauge theory this translates into the statement that $\vlim$ is lower than the speed of light in the absence of a medium, namely in the vacuum. The reason is that the absence of a medium in the gauge theory corresponds to the absence of a black hole on the gravity side, in which case $\vlim=1$ everywhere. 

Imagine now a heavy quark in the plasma. In the gravitational picture, this is described by a string that starts on the D-branes and falls through the horizon -- see fig.~\ref{cherenkov-with-quark}. In order to model a highly energetic quark we consider a string whose endpoint moves with an arbitrary velocity $v$ at an arbitrary radial position $\rho_0$. Roughly speaking, the interpretation of $\rho_0$ in the gauge theory is that of the inverse size of the gluon cloud that dresses the quark. This can be seen, for example, by holographically computing the profile of $\langle \mbox{Tr} F^2(x) \rangle$ around a static quark source dual to a string whose endpoint sits at $\rho=\rho_0$ \cite{size}.

\FIGURE{
\includegraphics[scale=.45]{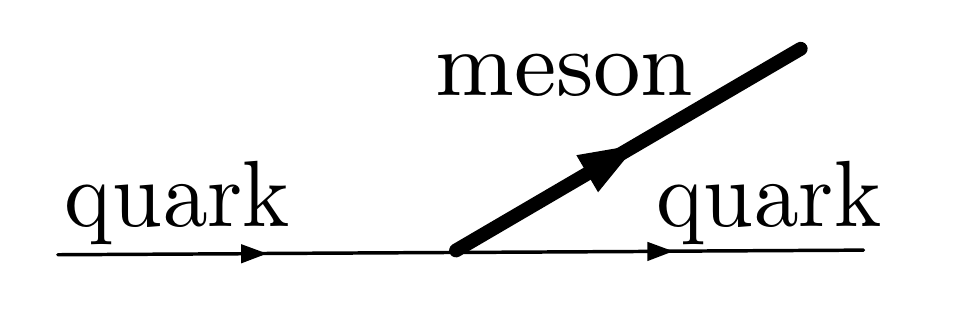}
\caption{Effective quark-meson coupling.} 
\label{quark-meson-coupling}} 
Two simple observations now lead to the effect that we are interested in. The first one is that the string endpoint is charged under the scalar and vector fields on the branes. In the gauge theory, this corresponds to an effective quark-meson coupling (see fig.~\ref{quark-meson-coupling}) of order $e \sim 1/\sqrt{\nc}$. We will derive these facts rigorously below, but physically they can be understood very simply. The fields on the branes describe fluctuations around the branes equilibrium configuration. The string endpoint pulls on the branes and therefore excites (i.e.~it is charged under) these fields. The branes tension is of order $1/g_s \sim \nc$, where $g_s$ is the string coupling constant, whereas the string tension is $\nc$-independent. This means that the deformation of the branes caused by the string is of order $e^2 \sim 1/\nc$. We thus conclude that the dynamics of the `branes+string endpoint' system is (a generalization of) that of classical electrodynamics in a medium in the presence of a fast-moving charge.

The second observation is that the velocity of the quark may exceed the limiting velocity of the mesons, since the redshift at the position of the string endpoint is smaller than at the bottom of the branes. As in ordinary electrodynamics, if this happens then the string endpoint loses energy by Cherenkov-radiating into the fields on the branes.\footnote{This can be viewed as a particular limit of string breaking -- see sec.~\ref{discussion}.}
 In the gauge theory, this translates into the quark losing energy by Cherenkov-radiating scalar and vector mesons. The rate of energy loss is set by the square of the coupling, and is therefore of order $1/\nc$.

\section{Quarks in the ${\cal N}=4$ SYM plasma}
\label{quarks}
The four-dimensional ${\cal N}=4$ SYM theory with gauge group $SU(\nc)$ at non-zero temperature is dual to type IIB string theory on the gravitational background sourced by $\nc$ black D3-branes. $\nf$ quark flavours may be introduced in the gauge theory by adding D7-branes on the gravity side. The relative orientation of the `colour' and `flavour' branes is summarised by the array
\begin{equation}
\begin{array}{ccccccccccc}
& 0 & 1 & 2 & 3 & 4& 5 & 6 & 7 & 8 & 9\\
\mbox{$\nc$ D3:} & \times & \times & \times & \times & & &  &  & & \\
\mbox{$\nf$ D7:} & \times & \times & \times & \times & \times  & \times & \times & \times &  &   \\
\end{array}
\label{D3D7}
\end{equation}
In the limit $\nf \ll \nc$ the backreaction of the D7-branes on the spacetime metric may be ignored and the D7-branes may be treated as probes in the gravitational background sourced by the D3-branes. Following \cite{prl,MMT2} we write the spacetime metric as $ds^2 = L^2 ds^2 (G)$, where 
\be
ds^2 (G) =  \frac{\rho^2}{2} \left[-\frac{f^2}{\tilde f} d t^2 + {\tilde f} dx_i^2\right]
+ \frac{1}{\rho^2} \left[ dr^2 + r^2 d \Omega^2_3 + dR^2 + R^2 d\vartheta^2 \right] \,,
\label{gst}
\ee
and 
\be
L^4=4\pi g_s \nc \ell_s^4 \sac \rho^2 = R^2 + r^2 \sac f = 1 - 1/\rho^4 \sac \tilde{f} = 1 + 1/\rho^4 \,.
\ee
The four gauge theory directions are $x^\mu=\{t, \vec x \}=\{t, x^i \}$, and they are identified with the 0123-directions shared by both sets of branes in \eqn{D3D7}. The metric inside the second set of brackets in \eqn{gst} is just the flat metric on $\mathbb{R}^6=\mathbb{R}^4 \times \mathbb{R}^2$, which corresponds to the 456789-directions in \eqn{D3D7}, written in terms of two sets of spherical coordinates $\{r, \Omega_3 \}$ and $\{R, \vartheta \}$. The coordinate $\rho$ is the overall radial coordinate in $\mathbb{R}^6$. This splitting is convenient since the D7-branes extend along the $\{r, \Omega_3 \}$-directions. All coordinates above are dimensionless, and they are related to their dimensionful counterparts (denoted with tildes) through 
\be
x^\mu = \pi T \, \tilde{x}^\mu \sac 
\{r, R, \rho \} = \frac{1}{\pi L^2 T} \, \{\tilde{r}, \tilde{R}, \tilde{\rho} \} \,.
\label{dimensionless}
\ee
In particular, this means that we are measuring energy and momentum in the gauge theory in units of $\pi T$. In addition, since the horizon of the metric \eqn{gst} in dimensionless coordinates lies at $\rho_\mt{hor}=1$, we see that the size of the horizon in physical units is proportional to the gauge theory temperature, i.e. $\tilde{\rho}_\mt{hor} \propto T$.
\newline
\newline
\noindent
{\it D7-brane embeddings}
\newline
\newline
We now specialize to $\nf=1$; we will discuss the case $\nf >1$ in sec.~\ref{discussion}.  The action governing the dynamics of a D7-brane in the background sourced by D3-branes takes the form 
\be
S_\mt{D7}= - T_\mt{D7} \int d^8 x \sqrt{-\det \left( g + 2\pi \ell_s^2 F \right)} + 
T_\mt{D7} \frac{\left( 2\pi \ell_s^2 \right)^2}{2} \int C_4 \wedge F \wedge F \,. 
\label{action}
\ee
In this equation $T_\mt{D7} = 1/g_s (2\pi)^7 \ell_s^8$ is the D7-brane tension, $\ell_s$ is the string length,  $x^a$ ($a=0, \ldots, 7$) are intrinsic coordinates on the brane's worldvolume, $g$ is the induced metric, $F=dA$ is the field-strength of the worldvolume $U(1)$ gauge field $A_a$, and $C_4$ stands for the pull-back of the spacetime Ramond-Ramond four-form potential sourced by the D3-branes. As we will see below, the term in the action involving $C_4$ will not contribute to any of our calculations. 

In order to describe the D7-brane embedding we use $x^a=\{ x^\mu, r, \Omega_3 \}$ as worldvolume coordinates. In other words, the brane extends along the gauge theory directions and the radial direction $r$, and it wraps an $S^3$ in the directions transverse to the D3-branes. Translational symmetry along $x^\mu$ and rotational symmetry along $\Omega_3$ then imply that the embedding must be specified as $R=R(r)$ and $\vartheta=\vartheta(r)$. Since $\vartheta$ is also a symmetry direction, a consistent solution is obtained by choosing $\vartheta=\mbox{const}$. A typical D7-brane embedding with different sets of coordinates suppressed is shown in fig.~\ref{embeddings}.
\FIGURE{
 \includegraphics[width= \textwidth]{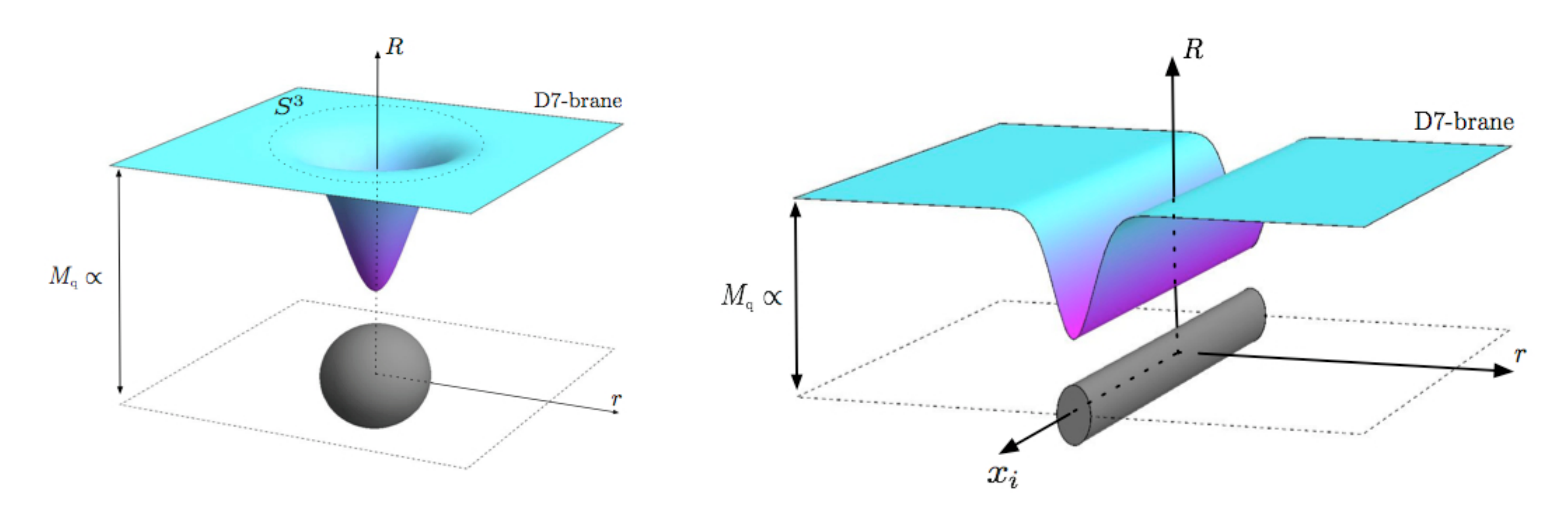} 
\caption{Minkowski-type D7-brane embedding showing the $S^3$ wrapped by the branes (left) and the gauge theory directions (right). The angular coordinate $\vartheta$ is suppressed in both cases. The asymptotic position of the brane is proportional to the quark mass $\mq$, whereas the size of the black hole horizon (shown in dark grey) is proportional to the temperature $T$.} 
\label{embeddings}}

Under these circumstances the induced metric on the D7-brane takes the form $ds^2 = L^2 ds^2 (g)$ with
\be
ds^2 (g) =  \frac{\rho^2}{2} \left[-\frac{f^2}{\tilde f} d t^2 + {\tilde f} d \vec x ^2\right]
+ \frac{(1+\dot R ^2)}{\rho^2}  dr^2+ \frac{r^2}{\rho^2} d \Omega^2_3 \,,
\label{metric}
\ee
where $\dot{R}=dR/dr$. The function $R(r)$ is determined by inserting \eqn{metric} in \eqn{action}, setting $F=0$, and varying with respect to $R(r)$. The resulting Euler-Lagrange equation of motion is
\be
\prt_r \left[ r^3 \left( 1 - \frac{1}{(r^2 + R^2)^4} \right) \frac{\dot R}{\sqrt{1 + {\dot R}^2}} \right] =
8 \frac{r^3 R}{(r^2 + R^2)^5} \sqrt{1 + {\dot R}^2} \,.
\label{embedeq}
\ee
In the limit $r \rightarrow \infty$, this equation leads to the asymptotic behaviour
\be
R(r) \simeq m + \frac{c}{r^2} + \cdots \,.
\label{asint}
\ee
Holography relates the dimensionless constants $m$ and $c$ to the quark mass and condensate as (see \cite{MMT2} for details) 
\be
\mq= \frac{1}{2}\sqrt{\lambda}\,T\,m \sac
\langle \bar{\psi} \psi \rangle =  -\frac{1}{8}\sqrt{\lambda}\,\nf\,\nc\,T^3\,c \,,
\ee
where $\lambda = \gym^2 \nc = 2\pi g_s \nc$ is the 't Hooft coupling.  An important point is that the constant $m$ can also be written in terms of the mass $\mmes$ of the lightest meson in the theory at zero temperature as \cite{MMT2}:
\be
m = \frac{2\mq}{\sqrt{\lambda}\, T} = \frac{\mmes}{2\pi T} \,.
\label{meson}
\ee

Eq.~\eqn{embedeq} cannot be solved analytically, but numerical solutions for any value of the asymptotic brane position, $m=R(r \rightarrow \infty)$, can be easily found. The constants $m$ and $c$ correspond to the two solutions at infinity of the second-order equation of motion \eqn{embedeq}. These solutions are mathematically independent, but not physically: once $m$ is specified, the requirement of regularity in the interior determines $c$. The physical solution is thus uniquely characterized by the value of $m$.\footnote{For thermodynamically stable embeddings. In the case of thermodynamically metastable or unstable embeddings, $c$ may be multivalued \cite{MMT,MMT2}.} In the gauge theory this translates into the statement that once the quark mass (and the temperature) are specified, the dynamics determines the quark condensate. 

Solutions of eq.~\eqn{embedeq} fall into two classes. For $m > 1.3$, i.e. for quark masses sufficiently larger than the temperature, the brane bends towards the horizon because of its gravitational pull, but the brane tension is able to compensate for this and the brane sits entirely outside the horizon, as in fig.~\ref{embeddings} and on the left-hand side of  fig.~\ref{transition}. 
\FIGURE{
\includegraphics[scale=.6]{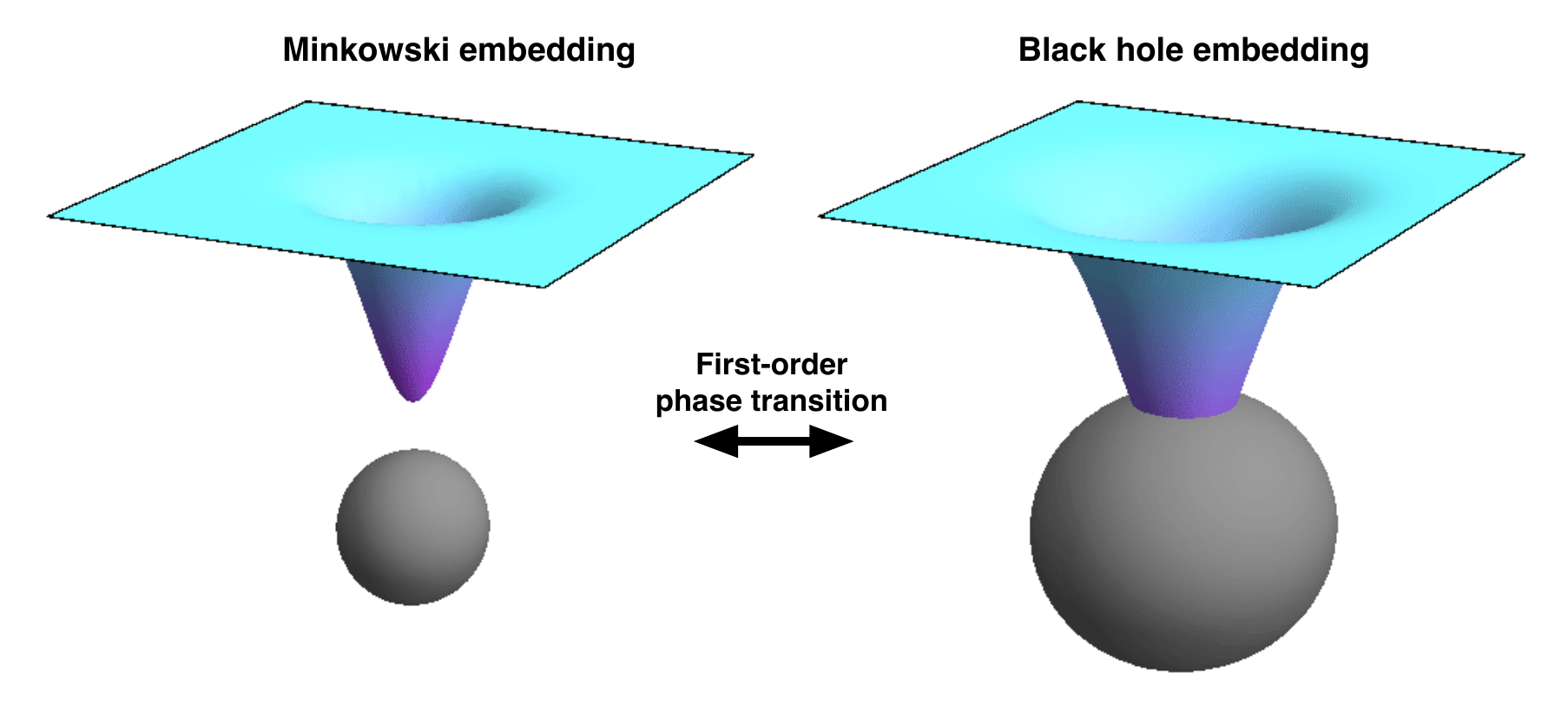}
\caption{First-order phase transition between Minkowski and black-hole type embeddings.} 
\label{transition}}
In this case the brane embedding is of the so-called `Minkowski' type \cite{MMT,MMT2}, and we will denote by $R_0=R(r=0)$ the radial position of the bottom of the branes. For Minkowski embeddings there is a one-to-one correspondence between $m$ and $R_0$, and the lowest value $m=1.3$ corresponds to $R_0=1.2$. 

In contrast, for $m<1.3$, the brane falls through the black hole horizon in a so-called `black-hole embedding', since in this case the induced metric on the branes possesses a horizon -- see fig.~\ref{transition}.  At $m=1.3$ a first-order phase transition between the two phases occurs, as indicated in fig.~\ref{transition}. As we will see below, in the Minkowski phase stable mesons exist, and their spectrum is discrete and gapped. In contrast, no stable mesons (in fact, no quasi-particles) exist in the black hole phase. For this reason, the phase transition above is interpreted in the gauge theory as a dissociation or `melting' phase transition for mesons in the $\caln =4$ plasma \cite{MMT,MMT2,melting}. 

The value $m=1.3$ thus corresponds to the (stable) Minkowski-type brane that comes closest to the horizon, and therefore to the one for which the in-medium meson dispersion relation is most dramatically modified with respect to that in the vacuum. For this reason, we have chosen the embedding with $m=1.3$ to illustrate some results in the sections below. Specifically, we see from eq.~\eqn{metric} that the local speed of light at the bottom of the brane is 
\be
\vlim =\left. \sqrt{-\frac{g_{00}}{g_{11}}}\right|_{r=0} =
\frac{f(R_0)}{\tilde{f}(R_0)}\,,
\label{vlim}
\ee
where we recall that $R_0=R(r=0)$. Since $m=1.3$ corresponds to $R_0=1.2$, the formula above gives $\vlim \simeq 0.35$, i.e.~in this case the limiting velocity of mesons in the plasma is about 1/3 of that in the vacuum.

\section{Meson dispersion relations  in the ${\cal N}=4$ SYM plasma}
\label{dispersionrelations}
Despite the fact that the ${\cal N}=4$ SYM theory is in a deconfined phase at any $T>0$, stable quark-antiquark states exist for sufficiently large $\mq/T$, and the spectrum of these mesons is discrete and gapped. In particular, scalar and vector mesons in the gauge theory are dual to regular, normalizable modes of the scalar and vector fields on the D7-brane. Here we will review the dispersion relations for these modes, which we will need in order to compute the quark energy loss below. The dispersion relation for (some) vector mesons in the D3/D7 system appeared in \cite{prl}, but no details were presented there. The dispersion relation for scalar mesons was first computed in \cite{MMT2} and then revisited in \cite{mit}. Here we will review the result in the geometric parametrization of \cite{mit}, which is particularly suited for calculating the energy radiated into these modes by the quark.

\subsection{Vector mesons}
These are dual to regular, normalizable fluctuations of the worldvolume gauge field $A$. 
The ${\cal N}=4$ SYM theory possesses an internal, global $SO(6)$ symmetry that is broken down to $SO(4)$ by the addition of quarks. In the string description, $SO(6)$ is the isometry group of the spacetime metric \eqn{gst}, whereas $SO(4)$ is the isometry group of the $S^3$ wrapped by the D7-branes. Under the preserved $SO(4)$ symmetry, meson modes decompose into singlet and non-singlet modes. Since we are interested in using the ${\cal N}=4$ SYM plasma as a toy model for the QCD plasma, and since QCD possesses no analog of the $SO(4)$ symmetry, we will focus on singlet modes.  The equation of motion for these modes receives no contribution from the second term in the action \eqn{action} \cite{spectroscopy}, and therefore we will ignore this term in the following. 

In conclusion, since we are interested in singlet modes, we only need to consider the first term in the action \eqn{action}. In addition, since we are only interested in their dispersion relation (as opposed to higher-order couplings), it suffices to expand this term to quadratic order in $F$ in the fixed worldvolume metric \eqn{metric}.\footnote{At higher orders gauge field fluctuations would mix with scalar fluctuations. Similarly, at higher orders singlet modes would generically mix with non-singlet modes.} The result is 
\be
S_\mt{vector} = - T_\mt{D7} L^4 \left(2 \pi l^2_s\right)^2 \int d^8 x \sqrt{-g}\, \frac{1}{4} F^{ab}F_{ab} \,,  
\label{vector}
\ee
leading to the equation of motion
\be
\sqrt{-g} \, \nabla_a F^{ab} = \prt_a(\sqrt{-g} \, F^{ab}) =0 \,.
\label{motion}
\ee
The metric $g$ that enters these expressions is that in eq.~\eqn{metric}, which contains no factors of $L$; these have been explicitly included in the prefactor of \eqn{vector}.

Singlet modes take the form 
\be
A_\mu = A_\mu (x^\mu, r) \sac A_r=A_r(x^\mu,r) \sac A_{\Omega_3} = 0 \,,
\label{zeromode}
\ee
i.e.~they have no components along the $S^3$ and depend only on $r$ and the gauge theory directions. The equations of motion are further simplified by the gauge choice $A_r=0$, which we will employ henceforth. In addition, we will work with the Fourier components of the gauge field defined through 
\be
A_\mu(t, x ,r) = 
\int  \frac{d\omega d^3 q}{{(2 \pi)}^4} \, A_\mu(\omega, q ,r) \, e^{-i \omega t + i q \cdot x} \,,
\label {fourier}
\ee
where $\omega$ and $q$ are the energy and the three-momentum of the meson, respectively. Finally, we choose the momentum to point along $x^1$ without loss of generality. 

Under the conditions above, the equations of motion for the transverse modes $A_2, A_3$ decouple from each other and from those for the longitudinal mode $A_0,A_1$, so we will study  them in turn. 
\newline
\newline
\noindent
{\it Transverse modes}
\newline
\newline
Let us collectively denote ${\cal A}=\{ A_2, A_3 \}$. Both modes obey identical equations of motion which take the form
\be
\partial_r \left( \sqrt{\-g} \, g^{rr} g^{33} \, \partial_r {\cal A} \right) - 
\sqrt{\-g} \, g^{33} \left( g^{00} \omega^2 + g^{11} q^2 \right) {\cal A} =0 \,.
\label{transg}
\ee
Upon using \eqn{metric} this becomes
\be
\partial_r \left( \frac{f r^3}{2\sqrt{1+  \dot{R}^2}} \, \partial_r {\cal A} \right)
+  \sqrt{1+ \dot{R}^2} \, \frac{r^3}{\rho^4} \left(\frac{\omega^2 \tilde{f}}{f}
- \frac{q^2 f}{\tilde{f}} \right) {\cal A} =0 \,.
\label{trans}
\ee
Since we are interested in regular, normalizable solutions, we can expand $\cala$ as
\be
\cala (\omega, q, r) = \sum_n \cala_n (\omega, q) \, \xi_n (q,r)
\label{expansion}
\ee
in terms of a basis of regular, normalizable eigenfunctions $\{ \xi_n (q,r) \}$ in the radial direction. These are solutions of eqn.~\eqn{trans} with $q$-dependent eigenvalues 
$\omega=\omega_n(q)$, i.e. they obey the eigenstate equation 
\be
- \partial_r \left( \frac{f r^3}{2\sqrt{1+  \dot{R}^2}} \, \partial_r \xi_n (q,r) \right) + 
\sqrt{1+ \dot{R}^2} \, \frac{f r^3}{\tilde{f} \rho^4}\, q^2 \, \xi_n (q,r) = 
\sqrt{1+ \dot{R}^2} \, \frac{\tilde{f} r^3}{f \rho^4}\, \omega_n(q)^2 \, \xi_n (q,r)
\label{eigenstate}
\ee
and satisfy the orthonormality relations 
\be
\int_0^\infty dr \frac{\tilde{f} r^3}{f \rho^4} \, \sqrt{1+\dot R^2}  \, 
\xi_m(q,r) \xi_n(q,r)  =\delta_{mn} \,.
\label{ortho}
\ee
As we will see in more detail below, the discreteness of the spectrum is guaranteed by the boundary conditions on the $\xi_n$: regularity at $r=0$ and normalizability at $r=\infty$. Inserting the expansion \eqn{expansion} in \eqn{trans}, and using the eigenstate equation \eqn{eigenstate} and the orthonormality relations \eqn{ortho}, we find that each of the $\cala_n(\omega,q)$ fields obeys an independent equation of the form
\be
\left[ \omega^2 - \omega_n^2(q) \right]  \cala_n(\omega,q) = 0 \,.
\label{fourdim}
\ee
Thus, through the expansion \eqn{expansion} we have `Kaluza-Klein-reduced' the five-dimensional field $\cala (\omega, q, r)$ to a discrete, infinite tower of independent four-dimensional fields $\{ \cala_n(\omega,q)\}$. Each of these fields is dual in the gauge theory to a transverse vector meson with dispersion relation $\omega=\omega_n(q)$, which is the physical meaning of the wave equation \eqn{fourdim}. In the gauge theory, each of the mesons in this infinite set is distinguished by its `internal' quantum number $n$. In the string description, each value of $n$ corresponds to a different, $q$-dependent radial `wave-function' $\xi_n (q,r)$. As we will see below, this structure of mesons in the fifth dimension will play an important role in determining the strength with which each of them couples to a quark. 
\FIGURE{
\begin{tabular}{cc}
\,\,\,\,\,\,\,\,\,\,\,\,
\includegraphics[width=0.40 \textwidth]{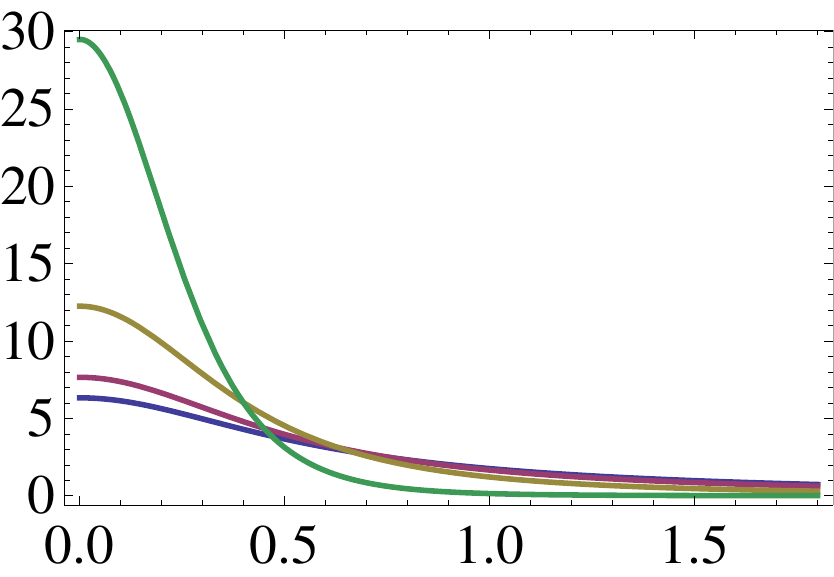}  
\put(-85,-9){$r$}
\put(-195,63){$\xi_0$}
&
\,\,\,\,\,\,\,\,\,\,\,\,\,\,\,\,\,\,\,\,\,
\includegraphics[width=0.40 \textwidth]{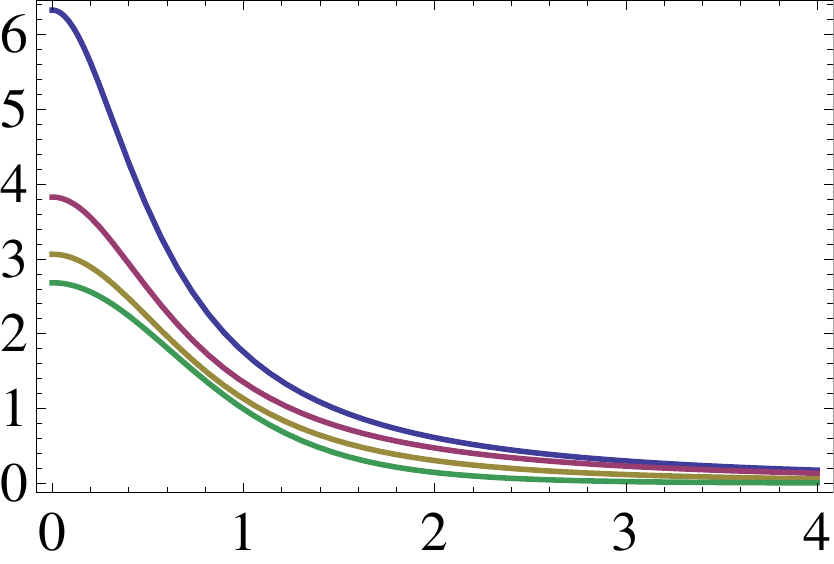}
\put(-93,-10){$\sqrt{q} \, r$}
\put(-204,63){$\xi_0/q$}
\vspace{7mm} \\
\,\,\,\,\,\,\,\,\,\,\,\,
\includegraphics[width=0.40 \textwidth]{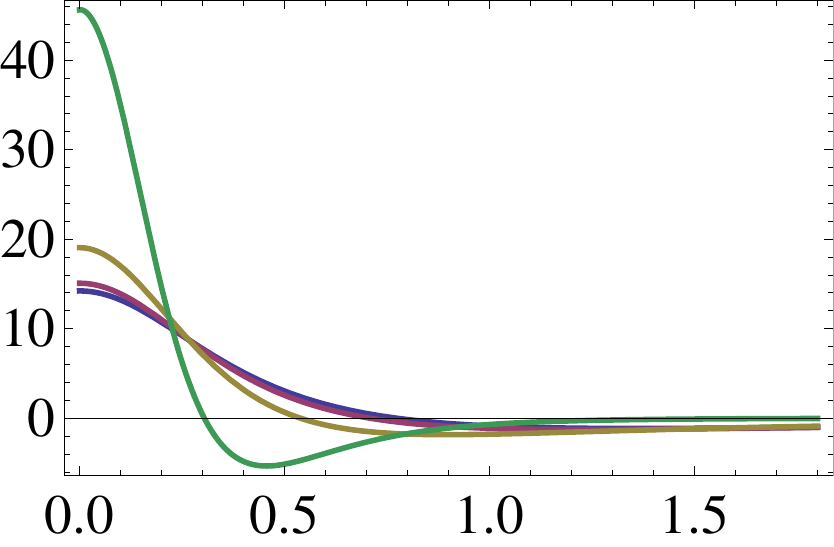}  
\put(-85,-9){$r$}
\put(-195,63){$\xi_1$}
&
\,\,\,\,\,\,\,\,\,\,\,\,\,\,\,\,\,\,\,\,\,
\includegraphics[width=0.42 \textwidth]{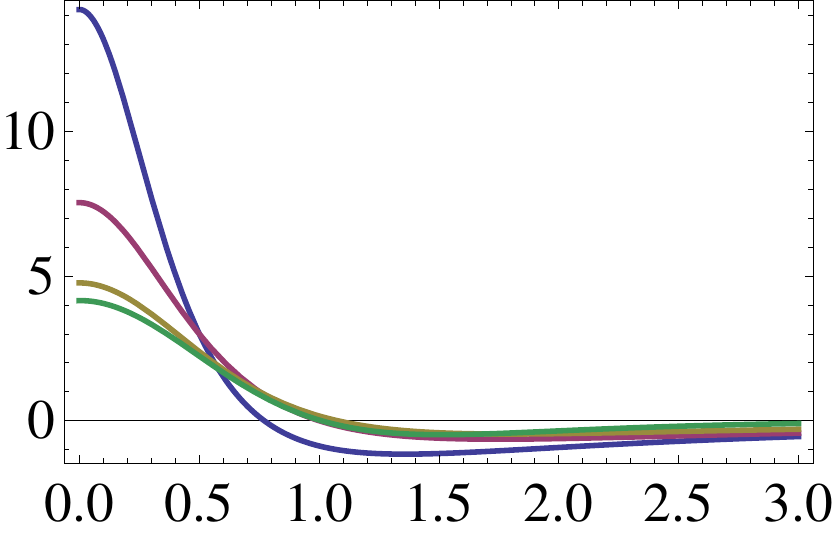}
\put(-95,-10){$\sqrt{q} \, r$}
\put(-210,63){$\xi_1/q$}
\vspace{7mm} \\
\,\,\,\,\,\,\,\,\,\,\,\,
\includegraphics[width=0.40 \textwidth]{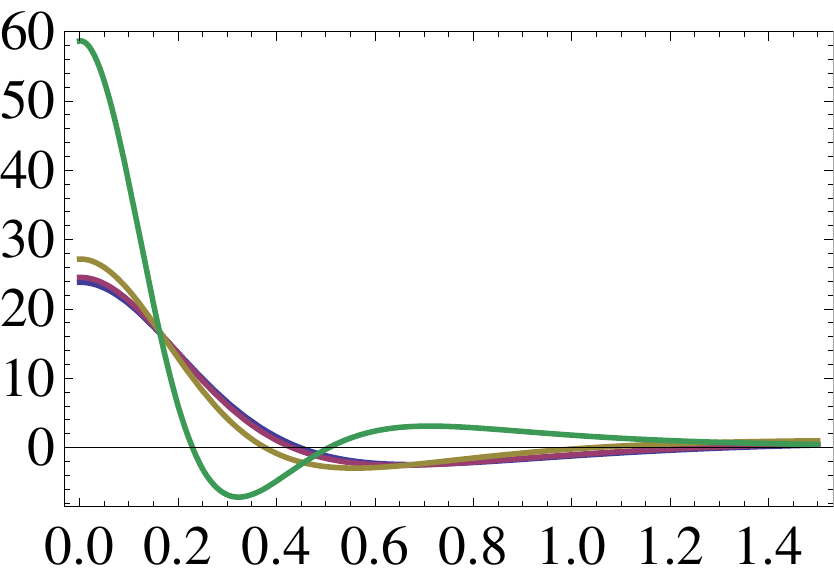}  
\put(-85,-10){$r$}
\put(-195,63){$\xi_2$}
&
\,\,\,\,\,\,\,\,\,\,\,\,\,\,\,\,\,\,\,\,\,
\includegraphics[width=0.42 \textwidth]{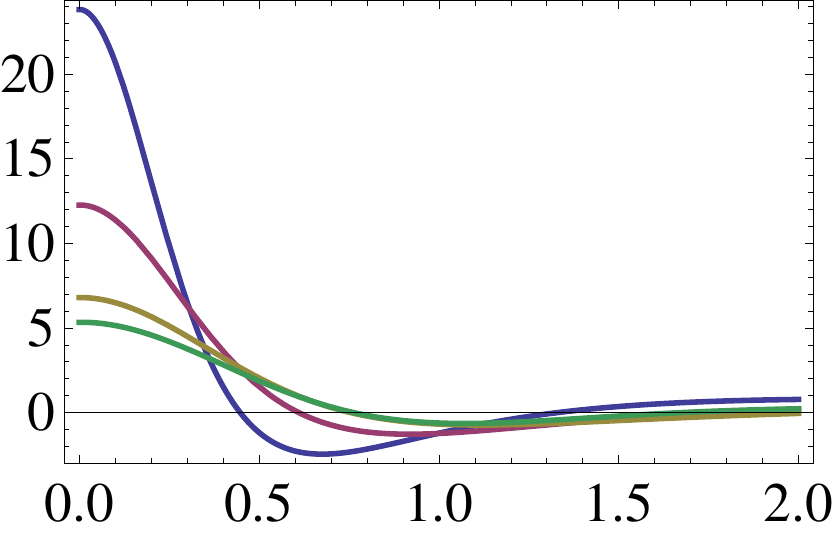}
\put(-98,-10){$\sqrt{q} \, r$}
\put(-212,63){$\xi_2/q$}
\end{tabular}
\caption{Normalized transverse vector meson radial profiles $\xi_{n} (q,r)$ for a D7-brane embedding with $m=1.3$. The blue, violet, brown and green curves (i.e.~bottom to top on the left, top to bottom on the right) correspond to $q=1,2,4, 11$, respectively. The rescalings on the right-hand side correspond to those in appendix \ref{hmwf}. Note that the area under the curves is not unity because of the non-trivial measure in eq.~\eqn{ortho}.} 
\label{transprofiles}}

Given that the brane embedding $R(r)$ entering eq.~\eqn{eigenstate} is only known numerically, the radial profiles must also be found numerically. The general solution of eq.~\eqn{eigenstate} behaves as $\xi_n  \sim a + b/r^2$ as $r \rightarrow 0$, and as $\xi_n  \sim \tilde{a} + \tilde{b}/r^2$ as $r \rightarrow \infty$, for some constants $a,b,\tilde{a},\tilde{b}$. Regularity at $r=0$ requires $b=0$, whereas normalizability imposes the condition $\tilde{a}=0$. For fixed $q$, these two requirements are mutually compatible only for a discrete set of values of the energy, $\omega_n(q)$. This is the origin of the dispersion relation. 
\FIGURE{
\,\,\,\,\,
\includegraphics[width=0.42 \textwidth]{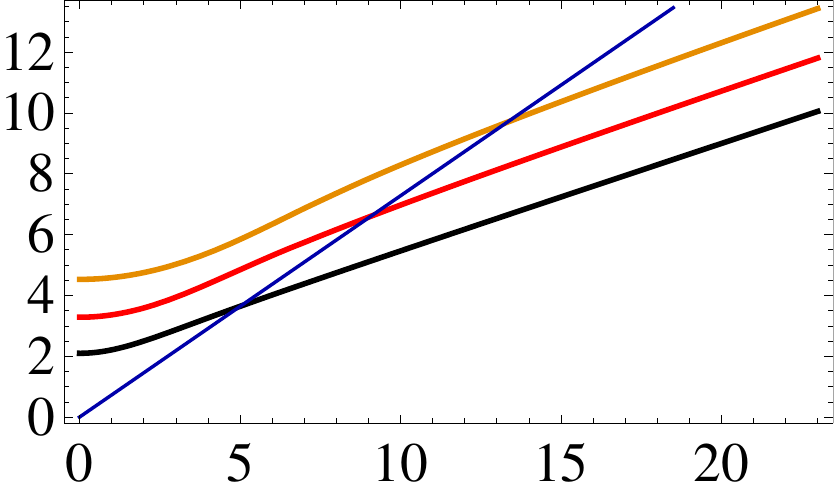}  
\put(-95,-12){$q/\pi T$}
\put(-200,50){\rotatebox{90}{{$\omega/\pi T$}}}
\vspace{3mm}
\caption{Dispersion relation for the first transverse modes $\xi_n$ of a heavy vector meson with $\vlim = 0.35$ in the ${\cal N}=4$ SYM plasma. The curves correspond to $n=0,1,2$ from bottom to top. The blue, continuous straight line corresponds to $\omega = v q$ with $\vlim<v<1$.} 
\label{transdispersion}}

Fig.~\ref{transprofiles} shows several numerically-obtained radial profiles of the first lowest-lying modes $\xi_n (q,r)$ for several values of $q$ for a D7-brane embedding with $m=1.3$. The corresponding values of the energy, $\omega_n (q)$, are given by the dispersion relation curves in figs.~\ref{dispersion} and \ref{transdispersion}. 
As is familiar with solutions of Schr\"odinger-like equations, the $n$th solution possesses  $n$ zeros. More importantly, we see that the radial wave-functions for all these modes become concentrated around the bottom of the brane, $r\simeq 0$, as 
$q \rightarrow \infty$. Relatedly, we observe that the limiting velocity of all these modes  agrees with the local speed of light at the bottom of the brane, eq.~\eqn{vlim}, as expected from the general argument in section \ref{universal}.
\newline
\newline
\noindent
{\it Longitudinal modes}
\newline
\newline
Eq.~\eqn{motion} with $b=0,1$ and $r$ yields two second-order dynamical equations and a first-order constraint equation, respectively, in which the longitudinal components $A_0$ and $A_1$ are coupled to one another. Only two out of the three equations are independent, which we take to be
\bea
\partial_r \left( \sqrt{-g} g^{rr} g^{00} \partial_r A_0  \right) +
iq \sqrt{-g} g^{11} g^{00} E &=& 0 \,, \label{modified} \\
i \omega g^{00} \partial_r A_0 - iq g^{11} \partial_r A_1 &=& 0 \,, \label{unmodified}
\eea
where we have introduced the gauge-invariant electric field
\be
E = F_{10} = iq A_0 + i \omega A_1 \,.
\ee
From these two equations, it is easy to see that $E$ satisfies 
\be
-i q  \, \partial_r \left( \frac{\sqrt{-g}\, g^{rr} g^{00} g^{11}}{q^2 g^{11}+w^2 g^{00}} \, \partial_r E \right) +
i q \sqrt{-g} \, g^{11}g^{00} E = 0 \,.
\label{Eequation}
\ee
In order to turn this into an eigenstate equation we introduce a new field $\Phi$ defined as 
\be
\Phi= \frac{\sqrt{-g} g^{rr} g^{00} g^{11}}{q^2 g^{11}+w^2 g^{00}} \, \partial_r E \,.
\label{phidef}
\ee
Eq.~\eqn{Eequation} then implies the inverse relation 
\be
\label{Eophi}
E=\frac{1}{\sqrt{-g}\, g^{11} g^{00}} \, \partial_r \Phi 
\ee
which, when substituted back into \eqn{phidef}, yields the equation of motion for 
$\Phi$: 
\be
 - \partial_r \left( \frac{1}{\sqrt{-g} \, g^{11} g^{00}} \, \partial_r \Phi \right) + 
 \frac{q^2 g^{11}+w^2 g^{00}}{\sqrt{-g} \, g^{rr} g^{00} g^{11}} \, \Phi = 0 \,.
 \label{PHIg}
\ee
Inserting the explicit form of the metric functions, we arrive at
\be
\partial_r \left( \frac{f \rho^4}{\tilde f r^3 \sqrt{1+\dot R^2}} \, \partial_r \Phi \right) + 2\frac{\sqrt{1+\dot R^2}}{f r^3} \left(w^2 -\frac{f^2}{\tilde f^2} q^2 \right) \Phi = 0 \,.
\label{PHI}
\ee
From this point onward, we proceed as in the case of transverse modes. We expand $\Phi$ as  
\be
\Phi(\omega,q,r) = \sum_n \Phi_n(\omega,q) \phi_n(q,r)
\label{phiexp}
\ee
in terms of a basis of regular, normalizable eigenfunctions $\{ \phi_n (q,r) \}$ in the radial direction. These are solutions of eq.~\eqn{PHI} with $q$-dependent eigenvalues $\omega=\omega_n(q)$, and are subject to the orthonormality relations 
\be
\int_0^\infty dr \, 2\frac{\sqrt{1+\dot R^2}}{f r^3} \phi_n(q,r) \phi_m(q,r)=\delta_{mn}\,.
\label{phiortho}
\ee
As in the case of transverse modes, the longitudinal modes $\Phi_n (\omega,q)$ obey the wave equation 
\be
\left[ \omega^2 - \omega_n^2(q) \right]  \Phi_n(\omega,q) = 0 \,,
\label{fourdimPhi}
\ee
as appropriate for a four-dimensional field with dispersion relation $\omega= \omega_n(q)$. Again, through the expansion \eqn{phiexp} we have Kaluza-Klein-reduced the five-dimensional field $\Phi (\omega, q, r)$ to a discrete, infinite tower of independent four-dimensional fields $\{ \varphi_n(\omega,q)\}$, each of which is dual to a longitudinal vector meson in the gauge theory.

The general solution of eq.~\eqn{PHI} behaves as $\phi_n  \sim a + b r^4$ as 
$r \rightarrow 0$, and as $\phi_n  \sim \tilde{a} + \tilde{b} \log r$ as $r \rightarrow \infty$, for some constants $a,b,\tilde{a},\tilde{b}$. Normalizability with respect to \eqn{phiortho} as 
$r \rightarrow 0$ requires that $a=0$, and regularity as $r\rightarrow \infty$ requires that 
$\tilde{b}=0$. As in the case of the transverse modes, for fixed $q$ these two requirements are mutually compatible only for a discrete set of energies $\omega_n(q)$.

Fig.~\ref{longprofiles} shows the electric field, eq.~\eqn{Eophi}, of several 
numerically-obtained radial profiles of the lowest-lying mode $\phi_{n=0}(q,r)$ for several values of $q$. The corresponding values of the energy, $\omega_{n=0}(q)$, are given by the dispersion relation curve in fig.~\ref{dispersion}. These results correspond again to a D7-brane embedding with asymptotic position $m=1.3$. We observe the same limiting velocity $\vlim=0.35$ given by the local speed of light at the bottom of the branes, eq.~\eqn{vlim}.  
\FIGURE{
\begin{tabular}{cc}
\,\,\,\,\,\,\,\,\,\,\,\,
\includegraphics[width=0.40 \textwidth]{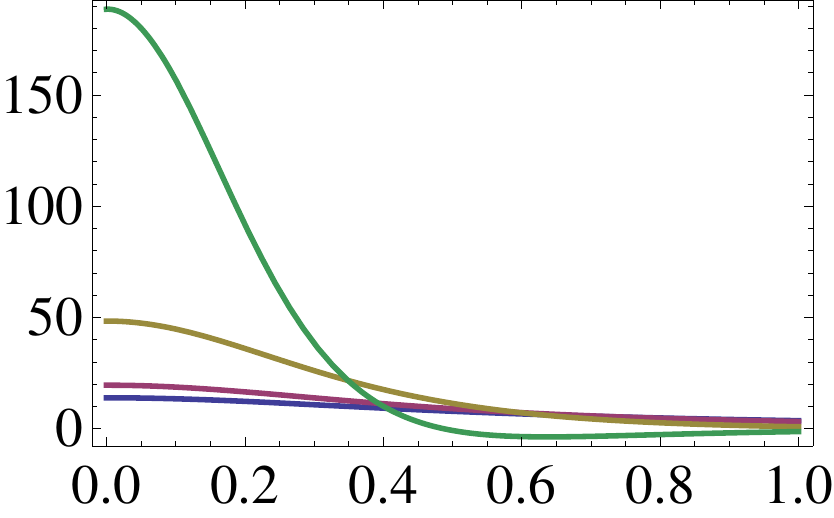}  
\put(-85,-9){$r$}
\put(-192,53){\rotatebox{90}{{$E_0$}}}
&
\,\,\,\,\,\,\,\,\,\,\,\,\,\,\,\,\,\,\,\,\,
\includegraphics[width=0.40 \textwidth]{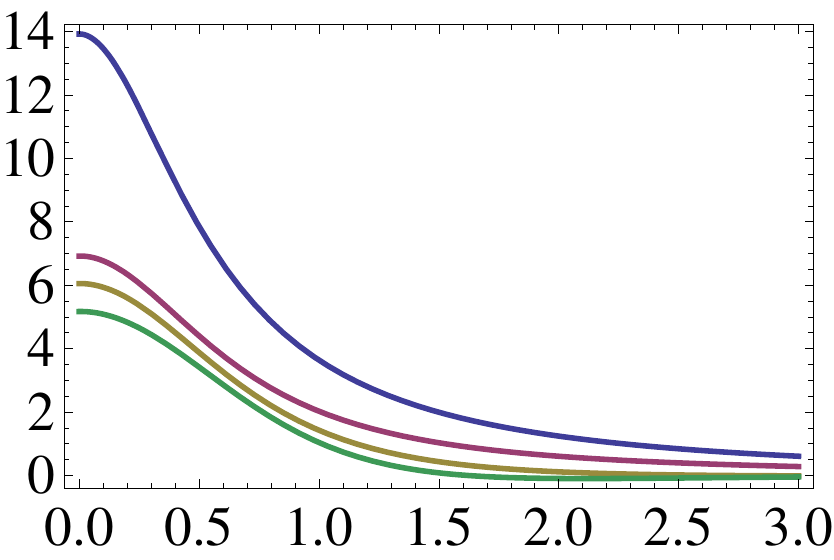}
\put(-95,-10){$\sqrt{q} \, r$}
\put(-194,48){\rotatebox{90}{{$E_0/q^{3/2}$}}}
\end{tabular}
\caption{Electric field $E_0(q,r)$ of the normalized longitudinal vector meson radial profiles $\phi_{0} (q,r)$ for a D7-brane embedding with $m=1.3$. The blue, violet, brown and green curves (i.e.~bottom to top on the left, top to bottom on the right) correspond to $q=1,2,4, 11$, respectively. The rescalings on the right-hand side correspond to those in appendix \ref{hmwf}. Note that the area under the curves is not unity because of the non-trivial measure in eq.~\eqn{phiortho}.} 
\label{longprofiles}}

\subsection{Scalar mesons}
\label{scalarsection}
The scalar fields on the brane get excited by the string endpoint because the string tension pulls on the brane. A crucial feature is the fact that the boundary conditions at the string endpoint imply that the string must end orthogonally on the brane. (The unfamiliar reader can find a concise derivation in appendix \ref{derivation}.) For this reason it is convenient to work with spacetime coordinates that locally parametrize the directions orthogonal to the brane. In the case of interest to us, the $\vartheta$ coordinate in \eqn{gst} satisfies this requirement, since the vector field $\partial/\partial \vartheta$ is orthogonal to the fiducial D7-brane embedding at each point on the brane. However, the $R$ coordinate does not meet this requirement, since $\partial/\partial R$ is in general not orthogonal to the brane due to the brane bending in the $r-R$ directions. We therefore follow \cite{mit} and work with two geometric coordinates $X^A$ defined as follows. At each point on the brane, the two-dimensional space orthogonal to the brane is spanned by the unit vectors 
\be
V_1 \propto \frac{\partial}{\partial R} - \dot{R}(r) \frac{\partial}{\partial r}
\sac V_2 \propto  \frac{\partial}{\partial \vartheta} \,,
\ee 
so a general vector orthogonal to the brane takes the form $U=X^A V_A$. For each vector $U$ we shoot off a geodesic with unit affine parameter that at the brane has $U$ as its tangent vector. The endpoint of this geodesic defines a point in a neighborhood of the brane. In this way we obtain a one-to-one correspondence (the so-called exponential map \cite{map}) between the values of $X^A$ and the points near the brane. In other words, $X^A$ are the coordinates orthogonal to the brane that we were seeking, since on the brane we have $\partial/\partial X^A = V_A$ by construction. In particular, this implies that in these coordinates $G_{AB}=\delta_{AB}$ when evaluated precisely on the brane, where $G$ is the spacetime metric \eqn{gst}. Note that in this section the ten dimensionless coordinates of spacetime are thus $\{ x^a, X^A \}$, with $a=0, \ldots , 7$ and $A=1,2$.

We chose the fiducial embedding of the brane to be given by $X^A=0$, so that the $X^A$ fields parametrize fluctuations around it. As shown in \cite{mit}, to quadratic order in these fields the D7-brane action takes the simple form
\be
S_\mt{scalar} = -T_\mt{D7} L^8 \int d^8 x \sqrt{-g} \left[ 
\frac{1}{2} g^{ab} \partial_a X^A \partial_b X^B G_{AB} + 
\frac{1}{2}  m_{AB}^2(x) X^A X^B  \right] \,,
\label{scalaraction}
\ee
where $g$ is the induced metric \eqn{metric} on the \emph{fiducial} embedding of the brane, i.e.~it is $X^A$-independent. As usual, this metric contains no factors of $L$, since this have been factored out explicitly in front of the action. The position-dependent mass matrix $m^2_{AB}(x)$ is diagonal and given in terms of geometric quantities as 
\bea
m_{11}^2 &=& R_{11} + R_{2112} + 2 R_{22} + \mbox{$^{(8)}R$} - R \,, \nn
m_{22}^2 &=& -R_{22} + R_{2112} \,, \label{M}
\eea
where
\bea
R_{2112} &=& V_2^{M} V_1^{N} V_1^{P} V_2^{Q} R_{MNPQ} \,, \\
R_{11} &=& V_1^M V_1^N R_{MN} \,, \\
R_{22} &=& V_2^M V_2^N R_{MN} \,.
\eea
$R_{MNPQ}$ and $R_{MN}$ are, respectively, the Riemann and the Ricci tensors of the ten-dimensional spacetime metric $G$, $R$ is the corresponding Ricci scalar, and $^{(8)}R$ is the Ricci scalar of the eight-dimensional induced metric on the brane $g$. Again none of these quantities contains any factors of $L$.

Using the fact that $G_{AB}=\delta_{AB}$ the action \eqn{scalaraction} leads to the equation of motion 
\be
\sqrt{-g} \, \nabla^2 X^A - \sqrt{-g} \, m^2 X^A =
\partial_a \left(  \sqrt{-g} \, \partial^a X^A  \right) -  \sqrt{-g} \, m^2 X^A = 0 \,,
\label{scalareom}
\ee
where $m=m_{11}$ or $m_{22}$ as appropriate. Following the vector meson case, we focus on the zero-mode of $X^A$ on the $S^3$, and work with its Fourier components $X^A(\omega, q, r)$, for which the equation of motion is
\be
\partial_r \left( \sqrt{-g} \, g^{rr} \partial_r X^A \right) - 
\sqrt{-g} \left( g^{00} \omega^2 + g^{11} q^2 + m^2 \right) X^A = 0 \,,
\label{Xg}
\ee
which upon substitution of the metric functions becomes
\be
\partial_r \left( \frac{f \tilde{f} r^3 \rho^2}{\sqrt{1+\dot R^2}}  \partial_r X^A \right) + 
f \tilde{f} r^3 \sqrt{1+\dot R^2} 
\left( \frac{2 \tilde{f}}{\rho^2 f^2} \omega^2 - 
\frac{2}{\rho^2 \tilde{f}} q^2  - m^2 \right) X^A =0 \,.
\label{X}
\ee

As usual, we expand $X^A$ as
\be
X^A (\omega, q, r) = \sum_n X_n^A (\omega, k) \, \varphi_n^A (k,r)
\label{Xexpansion}
\ee
in terms of a basis of normalizable eigenfunctions $\{ \varphi_n^A (k,r) \}$ in the radial direction. These are solutions of eq.~\eqn{X} with $q$-dependent eigenvalues 
$\omega=\omega_n^A(k)$, and are subject to the orthonormality relations 
\be
\int_0^\infty dr \,
\frac{2 \tilde{f}^2 r^3 \sqrt{1+\dot R^2}}{\rho^2 f}  \, 
\varphi_m^A (q,r) \varphi_n^A (q,r)  =\delta_{mn} \,.
\label{varphiortho}
\ee
Inserting the expansion \eqn{Xexpansion} in \eqn{X}, and using the orthonormality relations \eqn{varphiortho}, we find that each of the $X_n^A (\omega, q)$ fields obeys 
\be
\left[ \omega^2 - \omega_{nA}^2 (q) \right]  X_n^A (\omega,q) = 0 \,,
\label{waveX}
\ee
as expected. As explained in \cite{towards}, the modes $X^A_n (\omega,q)$ with $A=1,2$ correspond in the gauge theory to scalar and {\it pseudo}scalar mesons, respectively. 

Both masses (for $A=1,2$) in eq.~\eqn{X} behave as $m^2 \simeq -3 - m^2/r^2 + \cdots$ for $r\rightarrow \infty$ and $m^2 \simeq -c_1 + c_2 r^2 + \cdots$ for $r \rightarrow 0$, where $c_{1,2}$ are positive constants. It follows that the two independent solutions for 
$X^A$ behave as $1/r$ and $1/r^3$ for $r\rightarrow \infty$ and as $r^0$ and $1/r^2$ for $r \rightarrow 0$. Thus in this case normalizability requires that $X^A \sim 1/r^3$ for 
$r\rightarrow \infty$ and regularity requires that $X^A \sim r^0$ for $r \rightarrow 0$. As in the case of vector modes, for fixed $q$ these two requirements are compatible with each other only for a discrete set of energies $\omega_n^A(q)$.

Fig.~\ref{scalarprofiles} shows several numerically-obtained radial profiles of the lowest-lying 
mode $\varphi_{n=0}(q,r)$ for several values of $q$. The corresponding values of the energy, $\omega_{n=0}(q)$, are given by the dispersion relation curves in fig.~\ref{dispersion}. These results correspond again to a D7-brane embedding with asymptotic position $m=1.3$. We observe the same limiting velocity $\vlim=0.35$ given by the local speed of light at the bottom of the brane, eq.~\eqn{vlim}.  
\FIGURE[b!!!]{
\begin{tabular}{cc}
\,\,\,\,\,\,\,\,\,\,\,\,
\includegraphics[width=0.40 \textwidth]{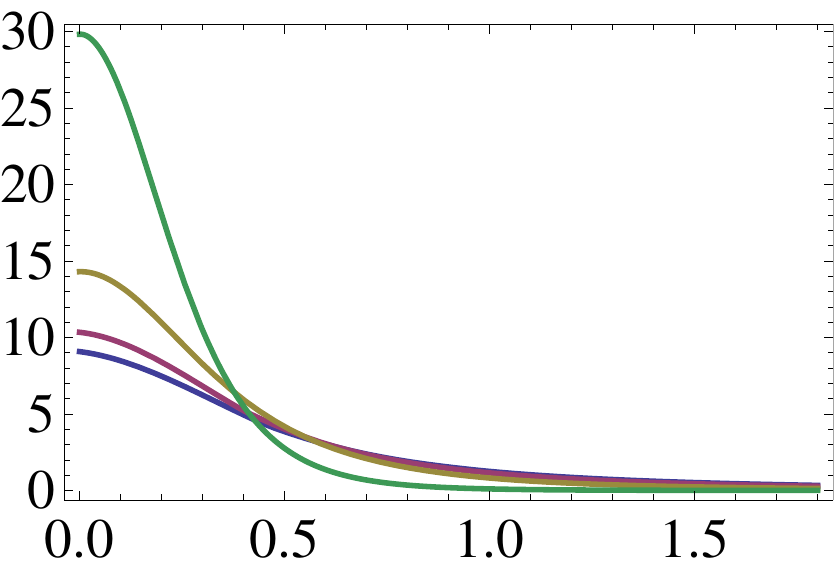}  
\put(-85,-9){$r$}
\put(-199,50){\rotatebox{90}{{$\varphi_0^{(A=1)}$}}}
&
\,\,\,\,\,\,\,\,\,\,\,\,\,\,\,\,\,\,\,\,\,
\includegraphics[width=0.40 \textwidth]{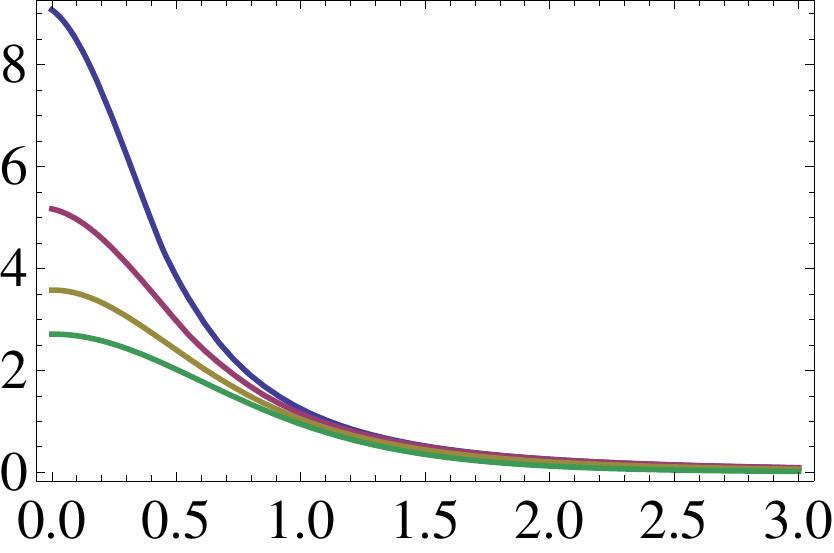}
\put(-95,-10){$\sqrt{q} \, r$}
\put(-199,40){\rotatebox{90}{{$\varphi_0^{(A=1)}/q$}}}
\vspace{7mm} \\
\,\,\,\,\,\,\,\,\,\,\,\,
\includegraphics[width=0.40 \textwidth]{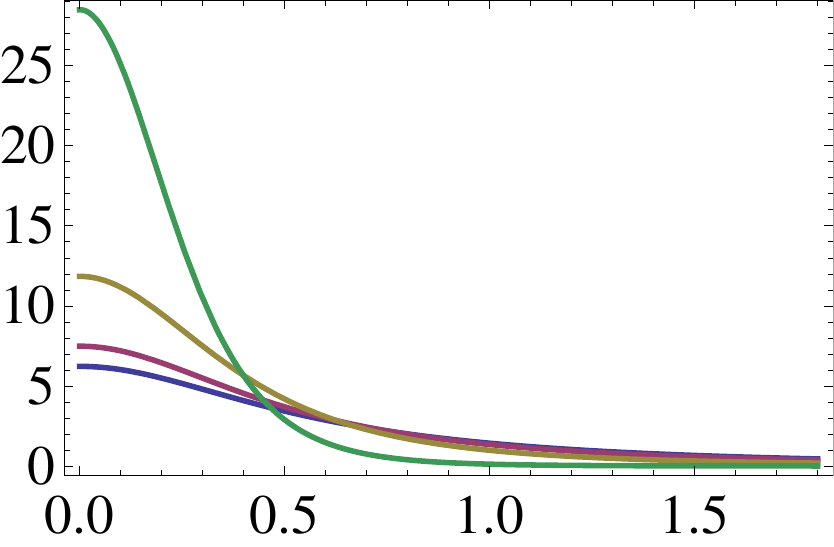}  
\put(-85,-9){$r$}
\put(-199,50){\rotatebox{90}{{$\varphi_0^{(A=2)}$}}}
&
\,\,\,\,\,\,\,\,\,\,\,\,\,\,\,\,\,\,\,\,\,
\includegraphics[width=0.40 \textwidth]{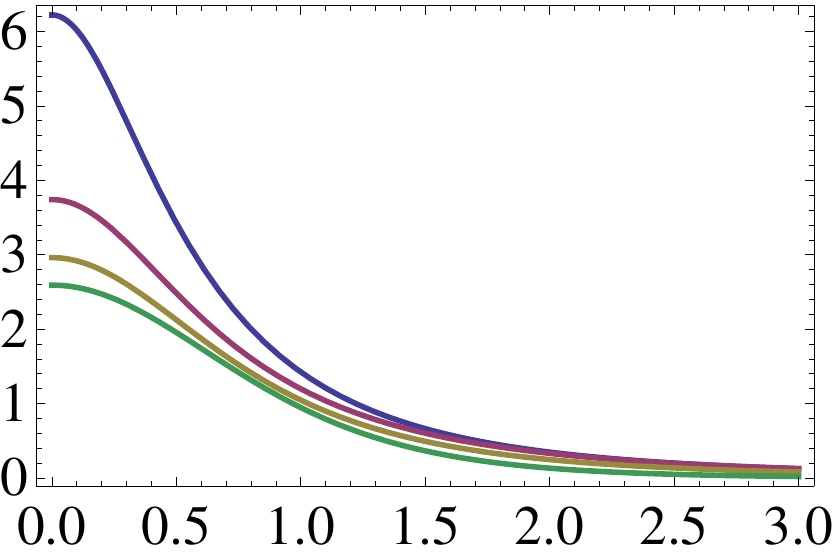}
\put(-95,-10){$\sqrt{q} \, r$}
\put(-199,40){\rotatebox{90}{{$\varphi_0^{(A=2)}/q$}}}
\end{tabular}
\caption{Normalized scalar (top plots) and pseudoscalar (bottom plots) meson radial profiles $\varphi_{n} (q,r)$ for a D7-brane embedding with $m=1.3$. The blue, violet, brown and green curves (i.e.~bottom to top on the left, top to bottom on the right) correspond to $q=1,2,4, 11$, respectively. The rescalings on the right-hand side correspond to those in appendix \ref{hmwf}. Note that the area under the curves is not unity because of the non-trivial measure in eq.~\eqn{varphiortho}.} 
\label{scalarprofiles}}

\section{Quark energy loss in the $\caln =4$ SYM plasma}
We now turn to the main topic of this paper, namely the rate at which a heavy quark traversing the ${\cal N}=4$ SYM plasma loses energy by Cherenkov-radiating mesons. As we will see in detail below, the quark acts a source for the brane stress-energy tensor, defined as  
\be
T_{ab} = - \frac{2}{\sqrt{-g}} \frac{\delta S_\mt{D7}}{\delta g^{ab}} \,.
\label{Tmunu}
\ee
This leads to its non-conservation, $\nabla^a T_{ab} \neq 0$, where $\nabla$ is the covariant derivative defined by the eight-dimensional worldvolume metric $g$. Under these circumstances the energy per unit time deposited on the brane by the quark is given by 
\be
\frac{dE}{dt} =- \int d^7 x \, \sqrt{-g} \, \nabla^a T_{a0} \,,
\label{formula}
\ee
where the integral is taken over the brane's worldspace. For the reader's convenience, a short derivation of this formula is provided in appendix \ref{loss}. Our task below will be to evaluate this formula for the cases of vector and scalar mesons. 

\subsection{Energy loss into vector mesons} 
The endpoint of an open string attached to the brane couples to the worldvolume gauge field, so the action \eqn{vector} is modified in the presence of the quark to
\be
S_\mt{vector} = 
- T_\mt{D7} L^4 \left(2 \pi l^2_s\right)^2 \int d^8 x \sqrt{-g}\, \frac{1}{4} F^{ab}F_{ab}  
-  \int d\tau A_a \frac{dx^a}{d\tau} \,,
\label{vectorquark}
\ee
where, as usual, the worldvolume metric $g$ contains no factors of $L$. The second term is the familiar coupling between an Abelian gauge field and a charged particle moving along a worldline parametrized as $x^a(\tau)$.\footnote{The relative normalization between the two terms in the action can be confirmed by noting that it ensures that supersymmetric BIon-like excitations on a D7-brane in flat space have tension $1/2\pi\ell_s^2$, as in \cite{bion}. \label{BIon}} In order to work with a canonically normalized gauge field we rescale $A \rightarrow e A$ with 
\be
e^2= \frac{1}{T_\mt{D7} \left(2 \pi l^2_s\right)^2 L^4} = \frac{8 \pi^4}{\nc} \,,
\label{e}
\ee
so that the action becomes 
\be
S_\mt{vector} = 
- \int d^8 x \sqrt{-g}\, \frac{1}{4} F^{ab}F_{ab}  - e \int d^8 x \, A_a J^a \,,
\label{vectorquarkaction}
\ee
where $J^a = \delta^{(7)}(x-x(\tau)) \, \dot{x}^a$. As anticipated in section \ref{universal}, the coupling $e$ between the quark and the vector mesons is of order $1/\sqrt{\nc}$, which justifies our neglect of terms of order higher than quadratic in the action. 

Using the definition \eqn{Tmunu}, the contribution from the gauge field to the brane's stress-energy tensor is easily calculated to be
\be
T_{ab} = F_{ac} F_b^{\,\,c} - \frac{1}{4} g_{ab} F^2 \,.
\label{Tvector}
\ee
In the presence of the string endpoint, the equation of motion \eqn{motion} for the gauge field  is modified to
\be
\sqrt{-g} \, \nabla_a F^{ab} = \prt_a(\sqrt{-g} \, F^{ab}) = e J^b\,,
\label{motionwithquark}
\ee
which implies the non-conservation of the stress-energy tensor
\be
\sqrt{-g} \, \nabla^a T_{ab} = e F_{ba} J^a \,.
\ee
Inserting this into the general formula \eqn{formula} yields the rate at which the quark deposits energy into the gauge field: 
\be
\frac{dE_\mt{vector}}{dt} = -e \int d^3 x dr d\Omega_3  \,   F_{0a} J^a \,.
\label{evec}
\ee
This formula has the simple interpretation of minus the work done on the quark by the gauge field. In order to evaluate it, we need to specify the quark trajectory. For simplicity, we will assume that the quark moves with constant velocity along a straight line at constant radial and angular positions, so we write
\be
J^a = \delta^{(3)}( \vec{x} - \vec{v} t ) \, \delta (r - r_0) \, \delta^{(3)}( \Omega - \Omega_0 ) \times (1, \vec{v}, 0, \vec{0} )\,.
\label{J}
\ee
In reality, $r_0$ and $v$ will of course decrease with time because of the black hole gravitational pull and the energy loss. However, we will concentrate on the initial part of the trajectory (which is long provided the initial quark energy is large) for which $r_0$ and $v$ are approximately constant \cite{trajectory} -- see fig.~\ref{cherenkov-with-quark}. The delta-functions in \eqn{J} allow us to perform the integral in \eqn{evec} and obtain 
\be
\frac{dE_\mt{vector}}{dt} =  - e v^i F_{0i}(t, \vec{v} t, r_0, \Omega_0) \,.
\label{vecloss}
\ee 

We thus see that we need to compute the electric field sourced by the string endpoint at the location of the string endpoint itself. To do so, we will solve the equation of motion \eqn{motionwithquark} by expanding the gauge field in normalizable modes in the radial direction, as in section \ref{dispersionrelations}. Note that the fact that the quark is localized on the $S^3$ means that it will radiate both into $S^3$ singlets and non-singlets. 
A simple group theory argument shows that these two types of contributions can be calculated separately and independently at the quadratic level at which we are working. For the reasons explained in section \ref{dispersionrelations}, we will only calculate the energy loss into singlet modes, whose form in Fourier-space we recall to be: 
\be
A_\mu = A_\mu (\omega, q, r) \sac A_r=0 \sac A_{\Omega_3} = 0 \,.
\ee
Without loss of generality, we choose $\vec{q}=(q,0,0)$ and $\vec{v}=(v\cos\theta, v\sin \theta, 0)$. After integrating over the $S^3$, the relevant Fourier-space components of the current are then
\be
J^\mu = 2\pi \delta( \omega - q v \cos \theta ) \, \delta (r - r_0) \times 
(1,v \cos \theta, v \sin \theta, 0) \,.
\label{Jmu}
\ee
We are now ready to compute the energy loss into the transverse and longitudinal modes of the gauge field. 
\newline
\newline
\noindent
{\it Transverse modes}
\newline
\newline
With the choice above the only transverse mode of the gauge field excited by the source is $\cala = A_2$, which couples to $\calj = J^2$. The equation of motion \eqn{trans} for this mode now becomes
\be
\partial_r \left( \frac{f r^3 \, \partial_r \cala}{2\sqrt{1+  \dot R ^2}}  \right)
+  \sqrt{1+\dot R^2} \frac{r^3}{\rho^4} \left(\frac{\omega^2 \tilde{f}}{f}
- \frac{q^2 f}{\tilde{f}} \right) \cala = {\tilde e} \calj \,, 
\label{transwith}
\ee
where $\tilde{e} = e/\Omega_3$ and the volume factor $\Omega_3=2\pi^2$ comes from integration over the $S^3$. We now follow section \ref{dispersionrelations} and solve \eqn{transwith} by expanding $\cala$ as in \eqn{expansion}, where the radial eigenfunctions $\{ \xi_n (q,r) \}$ satisfy exactly the same properties as in that section. In this case, inserting the expansion \eqn{expansion} in \eqn{transwith}, and using the eigenstate equation and the orthonormality relations, we find that eq.~\eqn{fourdim} becomes
\be
\left[ \omega^2 - \omega_n^2(q) \right]  \cala_n(\omega,q) 
= {\tilde e} \calj_n (\omega, q)\,,
\label{fourdimwith}
\ee
where
\be
\calj_n (\omega, q) = \int dr \calj(\omega, q) \xi_n (q,r) 
= 2\pi \delta\left(\omega- q v \cos\theta \right) v \sin \theta \, \xi_n (q,r_0) \,.
\label{Jn}
\ee
An important fact implied by eqs.~\eqn{fourdimwith}-\eqn{Jn} is that each of the four-dimensional meson modes $\cala_n(\omega,q)$ couples to the quark with an effective strength proportional to the value of  $\xi_n$ at the location of the quark:
\be
e_\mt{eff}(q,r_0) = e \, \xi_n (q,r_0) \,.
\label{eeff}
\ee
The intuition behind this is that the radial profiles 
$\xi_n(q,r)$ roughly play the role of a `wave function' in the fifth dimension for the corresponding meson mode $\cala_n(\omega,q)$. This fact will play 
an important role below.
 
With retarded boundary conditions, as appropriate for the reaction to the quark's passage, the solution of eq.~\eqn{fourdimwith} is 
\be
\cala_n(\omega,q) = 
\frac{\tilde{e} \calj_n (\omega, q)}{\left( \omega + i\epsilon \right)^2 - \omega_n^2(q)} \,.
\label{Asol}
\ee  
In order to evaluate the energy loss \eqn{vecloss},  we first express 
$F_{02}(t, \vec{v} t, r_0, \Omega_0)$ as an integral over its Fourier components:\footnote{Note that the singlet mode is independent of the $S^3$ position $\Omega_0$ of the string endpoint.}
\bea
\frac{dE_\mt{trans}}{dt} &=&  - e v^2 \partial_t A_2 (t, \vec{v} t, r_0, \Omega_0) \nn
&=& \int \frac{d\omega d^3q}{(2\pi)^4} (-e v \sin \theta) (-i\omega) \, \cala(\omega,q,r_0)\,
\left. e^{-i\omega t} e^{i q\cdot x} \right|_{\vec{x}=\vec{v}t} \,.
\eea 
Inserting the expansion \eqn{expansion} we obtain
\be
\frac{dE_\mt{trans}}{dt} = \sum_n \int \frac{d\omega d^3q}{(2\pi)^4} \, (e v \sin \theta)( i \omega) \,\cala_n(\omega,q) \xi_n(q,r_0) \, e^{-i\omega t} e^{i t q\cdot v} \,.
\ee
Substituting the solution \eqn{Asol} for $\cala_n$ and using the delta-function in \eqn{Jn} to integrate over frequencies, we arrive at
\be
\frac{dE_\mt{trans}}{dt} = \sum_n \int \frac{d^3q}{(2\pi)^3} \, 
(e v \sin \theta) ( i q v \cos \theta) \, 
\frac{\tilde{e} v\sin\theta}{\left( q v\cos\theta + i\epsilon \right)^2 - \omega_n^2(q)} \, \xi_n^2(q,r_0)\,.
\ee
Note that the two exponentials have cancelled out upon setting $\omega=q v \cos\theta$.
In order to integrate over momenta we set $d^3 q= 2\pi q^2 \,dq dz$, where 
$z=\cos \theta$, so that the integral above becomes
\be
\frac{dE_\mt{trans}}{dt} = \sum_n -\frac{e^2 v}{\Omega_3} \int_0^\infty \frac{dq}{2\pi} \,
q \, \xi_n^2 (q,r_0) \int_{-1}^1 \frac{dz}{2\pi i} 
\frac{z(1-z^2)}{(z+i\epsilon)^2 - z_n^2(q)} \,,
\label{zintegral}
\ee
where $z_n(q) = v_n(q)/v$, and $v_n (q) = \omega_n (q)/q$ is the phase velocity of the 
$n$-th mode. 
\FIGURE{
\includegraphics[scale=.70]{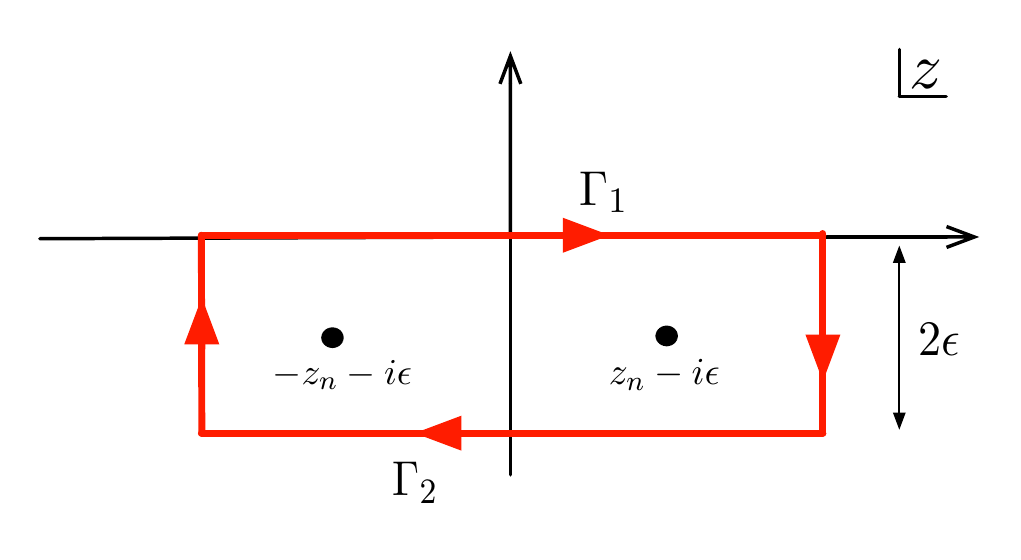}
\caption{Contour in the complex plane used to evaluate the $z$-integral in eq.~\eqn{zintegral}.} 
\label{contour}}
The integral over $z$ can be performed in the complex plane by considering the contour shown in fig.~\ref{contour}. The integral of interest corresponds to the integral over the segment $\Gamma_1$. In the limit $\epsilon \rightarrow 0$, this coincides with the integral over $\Gamma_2$, and the contribution from the vertical sides of the contour vanishes. Thus in this limit the integral over $z$ in \eqn{zintegral} equals 1/2 times the contour integral of fig.~\ref{contour}. Now consider the poles of the integrand, which lie at 
$z=\pm z_n -i\epsilon$. If $v < v_n(q)$ then $z_n(q) > 1$ and the poles lie outside the contour, so the integral vanishes. In contrast, if $v> v_n(q)$, then both poles lie inside the contour and they yield identical contributions equal to $(1-z_n^2)$. Taking into consideration the extra minus sign coming from the orientation of the contour, the final result is thus
\be
\frac{dE_\mt{trans}}{dt} = \sum_n \frac{e^2 v}{2\Omega_3} \int_0^\infty \frac{dq}{2\pi} \,
q \, \xi_n^2 (q,r_0)  \left( 1- \frac{v_n^2(q)}{v} \right)
 \Theta \left( 1 - \frac{v_n^2(q)}{v} \right) \,. 
 \label{transresult}
\ee
We see that the energy loss is a discrete sum over all mesons, as well as an integral over  all the momentum modes of each meson into which the quark is allowed to radiate. As expected for Cherenkov radiation, this can only happen if the velocity $v$ of the quark exceeds the phase velocity of the corresponding momentum mode, $v_n(q)$. For example, in the case of transverse vector mesons, the quark can only emit momentum modes to the right of the dashed, vertical line in  fig.~\ref{dispersion}. This cut-off is implemented by the Heaviside function in eq.~\eqn{transresult}.

Since the radial profiles $\xi_n(q,r_0)$ and the dispersion relations $v_n(q)$ entering 
eq.~\eqn{transresult} are only known numerically, the energy loss must also be evaluated numerically. The result for the $n=0$ term in the sum is plotted in fig.~\ref{result}.
As one may expect, for fixed $r_0$ the energy loss increases monotonically with $v$ up to the maximum allowed value of $v$, the local speed of light at $r_0$. In other words, a quark sitting at a fixed radial position radiates more the higher its velocity is. As $r_0$ decreases, the limiting velocity of the quark approaches that of the mesons from above. Therefore the quark and the meson dispersion relation curves cross at a higher momentum, i.e. the vertical dashed line in fig.~\ref{dispersion} moves to the right. This means that the characteristic momentum $q_\mt{char}$ of the modes contributing to the integral in \eqn{transresult} increases. As $r_0 \rightarrow 0$ these modes become increasingly peaked at small $r$ (see fig.~\ref{transprofiles}(left)), and their effective couplings to the quark $e_\mt{eff}(q_\mt{char},r_0)$ diverge. This explains why the energy loss at the maximum allowed value of the velocity diverges as $r_0 \rightarrow 0$. As we will discuss in section \ref{discussion}, however, this mathematical divergence is removed by physical effects that we have not taken into account. 
\FIGURE{
\begin{tabular}{ccc}
\,\,\,\,\,\,\,\,
\includegraphics[width=0.42 \textwidth]{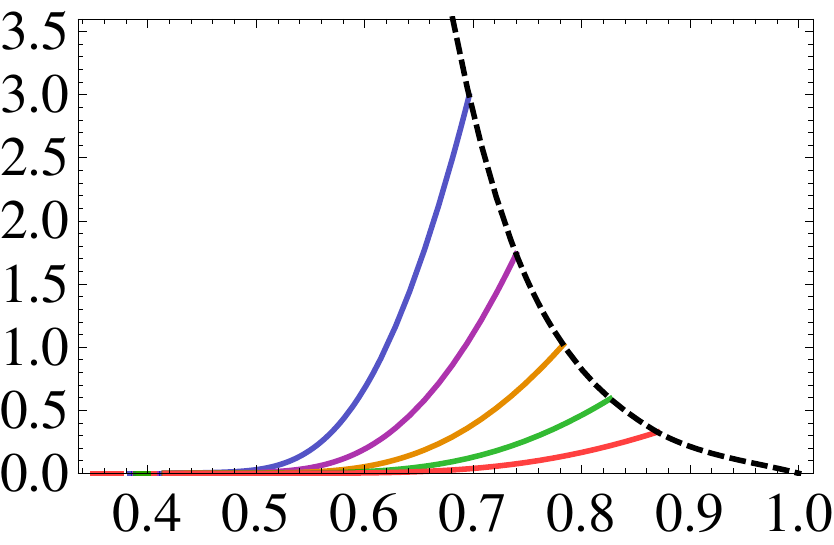} 
\put(-90,-10){$v$}
\put(-207,40){\rotatebox{90}{{$\frac{\nc}{\left( 2\pi T \right)^2} \frac{dE}{dt}$}}}
& \,\,\,\,\,\,\,\,\,\,\,\,\,&
\includegraphics[width=0.42 \textwidth]{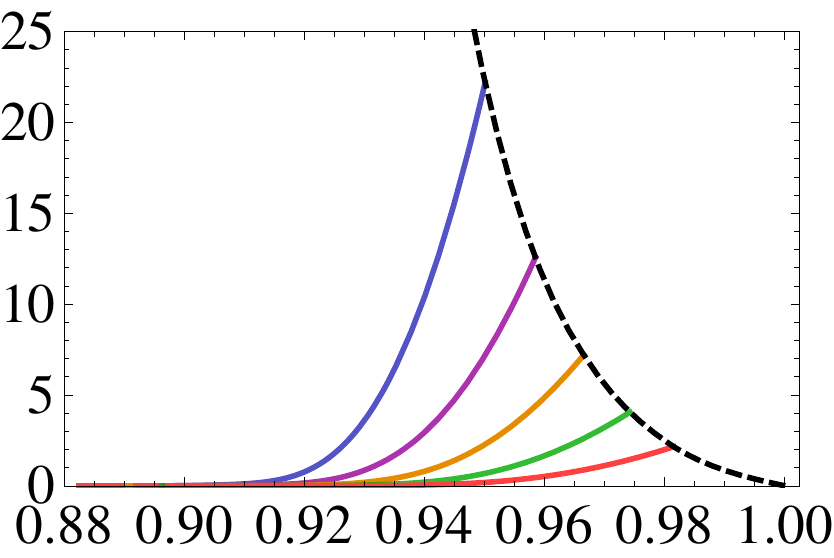} 
\put(-90,-10){$v$}
\put(-207,40){\rotatebox{90}{{$\frac{\nc}{\left( 2\pi T \right)^2} \frac{dE}{dt}$}}}
\\
\end{tabular}
\caption{Energy loss into the transverse vector mode ${\cal A}_{(n=0)}$ for an embedding with $m=1.32, R_0=1.2$ (left) and $m=2.0, R_0=2.0$ (right). The continuous curves correspond (from top to bottom) to $r_0=0.86, 0.97, 1.10, 1.25, 1.45$ (left) and to 
$r_0=1.50, 1.69, 1.91, 2.18, 2.52$ (right). The dotted curve is defined by the endpoints of the constant-$r_0$ curves.} 
\label{result}}
\newline
\newline
\noindent
{\it Longitudinal modes}
\newline
\newline
In the presence of the source eq.~\eqn{modified} becomes
\be
\partial_r \left( \sqrt{-g} g^{rr} g^{00} \partial_r A_0  \right) +
iq \sqrt{-g} g^{11} g^{00} E = \tilde{e} J^0 \,.
\ee
Eq.~\eqn{unmodified} remains unchanged, and together with the eq.~above it yields
\be
-i q  \, \partial_r \left( \frac{\sqrt{-g}\, g^{rr} g^{00} g^{11}}{q^2 g^{11}+w^2 g^{00}} \, \partial_r E \right) +
i q \sqrt{-g} \, g^{11}g^{00} E = \tilde{e} J^0 \,.
\label{EequationWith}
\ee
We now introduce a new field $\Phi$ defined as in eq.~\eqn{phidef}, but in this case the inverse relation \eqn{Eophi} is modified by the source:
\be
E=\frac{1}{\sqrt{-g}\, g^{11} g^{00}} \, \partial_r \Phi + 
\frac{1}{iq \, \sqrt{-g}\, g^{11} g^{00}} \, \tilde{e} J^0 \,.
\label{EophiWith}
\ee
Substituting back into eq.~\eqn{phidef} we find the new equation of motion for $\Phi$:
\be
 - \partial_r \left( \frac{1}{\sqrt{-g} \, g^{11} g^{00}} \, \partial_r \Phi \right) + 
 \frac{q^2 g^{11}+w^2 g^{00}}{\sqrt{-g} \, g^{rr} g^{00} g^{11}} \, \Phi = 
 \frac{1}{iq} \partial_r \left( \frac{1}{\sqrt{-g}\, g^{11} g^{00}} \, \tilde{e} J^0 \right) \,.
\ee
Inserting the explicit form of the metric functions, we finally arrive at
\be
\partial_r \left( \frac{f \rho^4}{\tilde f r^3 \sqrt{1+\dot R^2}} \, \partial_r \Phi \right) + 2\frac{\sqrt{1+\dot R^2}}{f r^3} \left(w^2 -\frac{f^2}{\tilde f^2} q^2 \right) \Phi = 
- \frac{1}{iq} \partial_r \left( \frac{f \rho^4}{\tilde{f} r^3 \sqrt{1+\dot R^2}} \, 
\tilde{e} J^0 \right) \,.
\label{PHIwith}
\ee

We now follow section \ref{dispersionrelations} and solve \eqn{PHIwith} by expanding 
$\Phi$ as in \eqn{phiexp}, where the radial eigenfunctions $\{ \phi_n (q,r) \}$ satisfy exactly the same properties as in that section. In this case, inserting the expansion \eqn{phiexp} in \eqn{PHIwith}, and using the eigenstate equation and the orthonormality relations, we find that eq.~\eqn{fourdimPhi} becomes
\be
\left[ \omega^2 - \omega_n^2(q) \right]  \Phi_n(\omega,q) 
= {\tilde e} J^0_n (\omega, q)\,,
\label{fourdimPhiwith}
\ee
whose solution with retarded boundary conditions is 
\be
\Phi_n(\omega,q) = 
\frac{\tilde{e} J^0_n (\omega, q)}{\left( \omega + i\epsilon \right)^2 - \omega_n^2(q)} \,.
\label{Phisol}
\ee  
The coefficients $J^0_n$ are given (after integration by parts) by
\be
J^0_n (\omega, q) = \int_0^\infty dr \frac{1}{iq} 
\frac{f \rho^4}{\tilde{f} r^3 \sqrt{1+\dot R^2}} \, J^0(\omega,q) \, \partial_r \phi_n (q,r) = 
\frac{2\pi}{iq} \delta(\omega- qv\cos \theta) \calf_n(q,r_0) \,,
\ee
with
\be
\calf_n(q,r) = - \frac{1}{\sqrt{-g}\, g^{11} g^{00}} \, \partial_r \phi_n (q,r) =
\frac{f \rho^4}{\tilde{f} r^3 \sqrt{1+\dot R^2}} \, \partial_r \phi_n (q,r)\,.
\ee
Note that the coefficients $\calf_n$ appear in the expansion of the electric field \eqn{EophiWith}, i.e.
\be
E(\omega,q,r) = -\sum_n  \Phi_n(\omega,q) \calf_n(q,r)
+ \frac{1}{iq \, \sqrt{-g}\, g^{11} g^{00}} \, \tilde{e} J^0 (\omega,q,r)\,.
\label{ignore}
\ee
As in the case of transverse modes, we see from eq.~\eqn{fourdimPhiwith} that the effective coupling between a longitudinal meson $\Phi_n(\omega,q)$ and the quark is determined by the radial wave function of the meson, in this case 
\be
e_\mt{eff}(q,r_0)=e \, \calf_n (q,r_0) \,.
\label{eeffLong}
\ee

Our task now is to compute the rate of energy loss into longitudinal meson modes. For this purpose, eq.~\eqn{vecloss} instructs us to evaluate the electric field at the location of the quark. If we naively do so using the expression \eqn{EophiWith} for $E$ then the second term gives a divergent result, since $J^0(r_0) \propto \delta(0)$. However, this divergence is unphysical: if one replaces the delta-function by a smooth charge distribution, then the integral over space in \eqn{evec} vanishes. Indeed, suppose that the current \eqn{Jmu} is replaced by 
\be
J^a = \varrho^{(3)}( \vec{x} - \vec{v} t, r, \Omega )  \times (1, \vec{v}, 0, \vec{0} )\,,
\label{Jreg}
\ee
where $\varrho$ is a smooth function. Then in Fourier space
\be
J^a = 2\pi \, \delta(\omega-\vec{q} \cdot \vec{v}) \, \varrho^{(3)}( \vec{q}, r, \Omega )  \times (1, \vec{v}, 0, \vec{0} )
\label{JregFourier}
\ee
and the energy loss \eqn{evec} is 
\bea
\frac{dE}{dt} &=& - e\int dr d\Omega \int d^3 x \,  E(t,\vec{x},r,\Omega) \cdot
J(t,\vec{x},r,\Omega) \nn
&=& - e\int dr d\Omega \int \frac{d\omega d\tilde{\omega}}{(2\pi)^2} 
e^{-i \omega t -i\tilde{\omega} t} \int \frac{d^3 q}{(2\pi)^3} z \,
E(\tilde{\omega}, -\vec{q}, r, \Omega) \, J^1 (\omega,\vec{q},r,\Omega) \,,
\eea
where as usual $z=\cos \theta$ is the relative angle between $\vec{q}$ and $\vec{v}$ and $d^3q = 2\pi q^2 \, dq dz$. If we substitute the term in $E$ that is proportional to $J^0$ we see that the integrand is proportional to 
\be
\frac{1}{iq} \, \delta(\tilde{\omega} + q v z)\, \delta(\omega - qvz) \, \varrho(-\vec{q},r,\Omega) \, \varrho(\vec{q},r,\Omega) \, z \,.
\ee
This is odd under $\vec{q} \rightarrow - \vec{q}$ (since $\omega$ and $\tilde{\omega}$ are dummy variables) and therefore the integral over $z$ vanishes. 

We therefore conclude that we can neglect the second term in \eqn{ignore} in order to evaluate \eqn{vecloss}. Following the previous section we have
\bea
\frac{dE_\mt{long}}{dt} &=& e v^1 E(t,\vec{v} t, r_0, \Omega_0) \nn
&=& \sum_n - \int \frac{d\omega d^3q}{(2\pi)^4} \left( e v \cos\theta \right) 
\Phi_n (\omega,q) \calf_n(q,r_0)  \, e^{-i\omega t} e^{i t q\cdot v} \nn
&=& \sum_n - \int \frac{d^3q}{(2\pi)^3} \left( e v \cos\theta \right) 
\frac{\tilde{e}}{iq} \,
\frac{1}{\left( q v\cos\theta + i\epsilon \right)^2 - \omega_n^2(q)} \,
\calf_n^2(q,r_0) \nn
&=& \sum_n -\frac{e^2}{\Omega_3  v} \int_0^\infty \frac{dq}{2\pi} \,
\frac{1}{q} \, \calf_n^2 (q,r_0) \int_{-1}^1 \frac{dz}{2\pi i} 
\frac{z}{(z+i\epsilon)^2 - z_n^2(q)} \,.
\eea
The $z$-integral can be evaluated using the same contour of fig.~\ref{contour}. In this case each pole contributes -1/2, so the final result is 
\be
\label{dEdtVlong}
\frac{dE_\mt{long}}{dt} =  \sum_n \frac{e^2}{2 \Omega_3  v} 
\int_0^\infty \frac{dq}{2\pi} \, \frac{1}{q} \, \calf_n^2 (q,r_0) \,
\Theta \left( 1 - \frac{v_n^2(q)}{v} \right) \,.
\ee

The energy loss into the $n=0$ mode is shown in fig.~\ref{result-long}. The same comments as in the case of transverse modes apply here.
\FIGURE{
\begin{tabular}{ccc}
\,\,\,\,\,\,\,\,
\includegraphics[width=0.42 \textwidth]{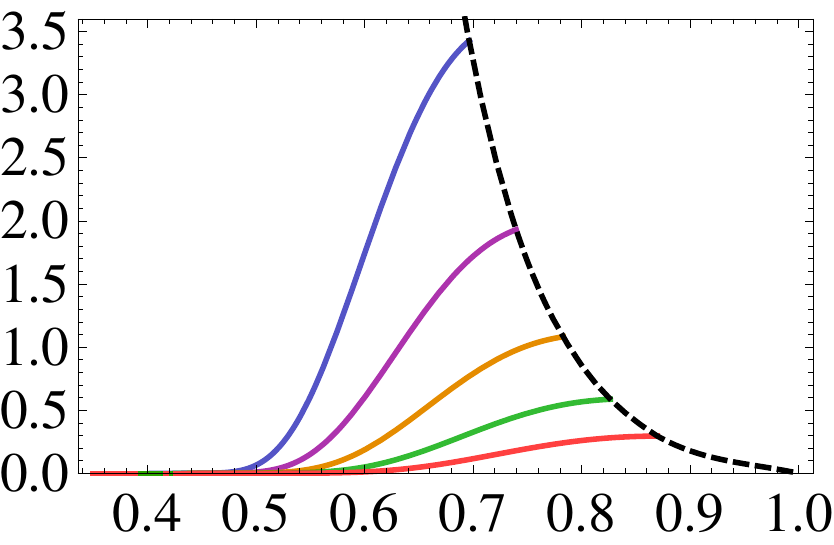} 
\put(-90,-10){$v$}
\put(-207,40){\rotatebox{90}{{$\frac{\nc}{\left( 2\pi T \right)^2} \frac{dE}{dt}$}}}
& \,\,\,\,\,\,\,\,\,\,\,\,\,&
\includegraphics[width=0.42 \textwidth]{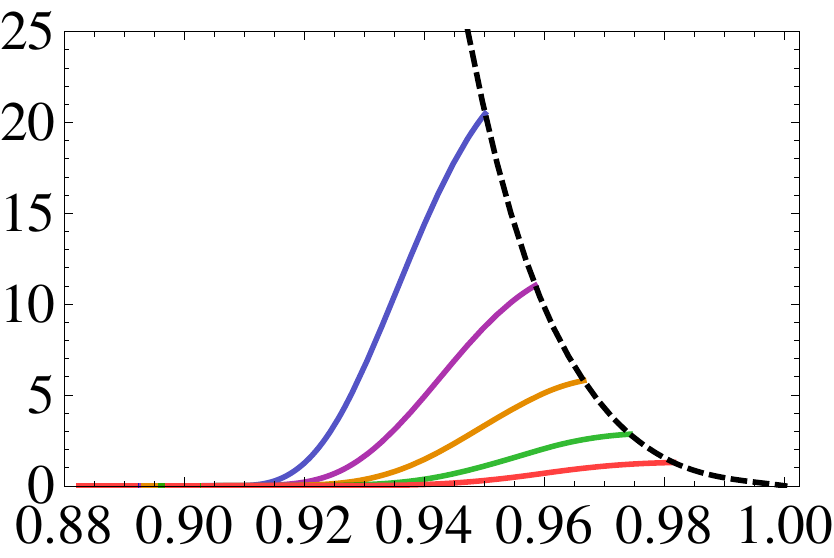} 
\put(-90,-10){$v$}
\put(-207,40){\rotatebox{90}{{$\frac{\nc}{\left( 2\pi T \right)^2} \frac{dE}{dt}$}}}
\\
\end{tabular}
\caption{Energy loss into the longitudinal vector mode $\Phi_{(n=0)}$ for an embedding with $m=1.32, R_0=1.2$ (left) and $m=2.0, R_0=2.0$ (right). The continuous curves correspond (from top to bottom) to $r_0=0.86, 0.97, 1.10, 1.25, 1.45$ (left) and to 
$r_0=1.50, 1.69, 1.91, 2.18, 2.52$ (right). The dotted curve is defined by the endpoints of the constant-$r_0$ curves.} 
\label{result-long}}

\subsection{Energy loss into scalar mesons} 
The endpoint of an open string attached to the brane couples to the worldvolume scalar fields, so in the presence of the quark the action \eqn{scalaraction} is modified to
\bea
S_\mt{scalar} &=& -T_\mt{D7} L^8 \int d^8 x \sqrt{-g} \left[ 
\frac{1}{2} g^{ab} \partial_a X^A \partial_b X^B G_{AB} + 
\frac{1}{2}  m_{AB}^2(x) X^A X^B  \right] \nn
&& - \frac{L^2}{2\pi \ell_s^2}  \int d\tau \sqrt{- \dot{x}^2} \, G_{AB} \, X^A n^B \,,
\label{scalaractionWith}
\eea
where $ \dot{x}^2 = g_{ab}\dot{x}^a \dot{x}^b$ and $n^A$ is the unit vector that is tangent to the string and orthogonal to the brane at the string endpoint. As usual, in the last term we have explicitly factored out  the $L$-dependence associated to the spacetime metric \eqn{gst}. The square-root factor in the last term is necessary to make the integrand a scalar density on the worldline of the string endpoint, which is contained in the brane's worldvolume. We emphasize that the indices of $\dot{x}^a$ are not contracted with the gauge theory metric $\eta_{ab}$ but with the D7-brane metric $g_{ab}$. The relative normalization between the two lines in \eqn{scalaractionWith} can be confirmed as described in footnote \ref{BIon}. As explained in appendix \ref{derivation}, the boundary condition at the string endpoint implies that the string ends orthogonally on the brane. The unit normal in the last term of the action \eqn{scalaractionWith} means that the string couples to the scalar that parametrizes the direction along which the string pulls on the brane, as one may intuitively expect.

In order to work with canonically normalized fields we rescale 
$X^A \rightarrow \sqrt{T_\mt{D7} L^8} \, X^A$ so that the action becomes
\be
S_\mt{scalar} = - \int d^8 x \sqrt{-g} \left[ 
\frac{1}{2} g^{ab} \partial_a X^A \partial_b X^B G_{AB} + 
\frac{1}{2}  m_{AB}^2(x) X^A X^B  \right]  
-  e \int d^8 x \, J_A  X^A \,,
\label{scalaractionWithBis}
\ee
where $J_A =  \delta^{(7)}(x-x(\tau)) \, \sqrt{- \dot{x}^2} \, G_{AB} \, n^B$ and $e$ is the same coupling constant defined in \eqn{e}. Since the scalars do not interact with each other, they give independent contributions to the stress tensor of the brane. Using the definition \eqn{Tmunu}, the contribution from either scalar field is easily found to be
\be
T_{ab} = \nabla_a X \nabla_b X - 
\frac{1}{2} g_{ab} \left[ \left(\nabla X \right)^2  + m^2 X^2 \right] +
X^2 \, \frac{\delta m^2}{\delta g^{ab}} \,.
\ee
For ease of notation, in this equation we have dropped the superindex `A' on the scalar, and we will continue to do so below. The last term originates from the non-trivial dependence of the scalar masses \eqn{M} on the metric. Fortunately, we will see that we do not need to evaluate this term explicitly in order to compute the divergence of the stress-tensor. 

In the presence of the string endpoint, the equation of motion \eqn{scalareom} for the scalars is modified to 
\be
\sqrt{-g} \left( \nabla^2 -  m^2 \right) X = e J \,.
\label{scalareomWith}
\ee
Using this, the divergence of the stress tensor takes the form
\be
\sqrt{-g} \, \nabla^a T_{ab} = J \, \nabla_b X -X^2 \, \nabla_b m + 
\nabla^a \left( X^2 \, \frac{\delta m^2}{\delta g^{ab}} \right) \,.
\ee
The second term on the right-hand side is due to the possible spacetime dependence of the scalar masses $m(x)$, but it vanishes identically for the case of interest here, $b=0$, because of the time-translation invariance of the theory. The last term on the right-hand side vanishes when evaluated on a solution of the equations of motion, even in the presence of the source $J$. To see this, recall that the stress tensor must be identically conserved in the absence of the source because of the diffeomorphism invariance of the brane's worldvolume theory. This means, in particular, that when $J=0$ we have
\be
X \, \nabla^a \left( \frac{\delta m^2}{\delta g^{a0}} \right) + 
2  \left( \frac{\delta m^2}{\delta g^{a0}} \right) \nabla^a X = 0 \,.
\label{vanish}
\ee
The key point now is that this equation is linear in $X$. Since the solution in the presence of the source is a linear supersposition of solutions of the source-less equation, linearity of \eqn{vanish} implies that this expression also vanishes for solutions of 
eq.~\eqn{scalareomWith} with $J\neq0$.  We thus conclude that (the time component of) the non-conservation of the stress-tensor in the scalar sector takes the form
\be
\sqrt{-g} \, \nabla^a T_{a0} = J \, \nabla_0 X \,.
\label{non}
\ee

As in the case of vector mesons, we consider a rectilinear quark motion with constant velocity, in which case 
\be
J =  \sqrt{-\dot{x}^2 (r_0)}  \, 
\delta^{(3)}( \vec{x} - \vec{v} t ) \, \delta (r - r_0) \, \delta^{(3)}( \Omega - \Omega_0 ) \,.
\ee
Note that, although the velocity $v$ is the quark velocity as seen by a gauge theory observer, the prefactor above is not just $\sqrt{1-v^2}$ but depends non-trivially on the quark position in the radial direction through
\be
 \sqrt{-\dot{x}^2 (r_0)}  = \sqrt{-g_{00} (r_0) - g_{11}(r_0) v^2} \,.
\label{sqrt}
\ee
Because of the black hole redshift, for fixed $r_0$ this factor vanishes before $v$ reaches the speed of light, i.e.~at $v<1$. As we will see, this fact is responsible for a qualitative difference between the energy radiated into scalar and into vector mesons. 

Following section \ref{scalarsection}, we focus on the zero-mode of $X$ on the $S^3$, and work with its Fourier components $X(\omega, q, r)$, for which the equation of motion takes the form
\be
\partial_r \left( \sqrt{-g} \, g^{rr} \partial_r X \right) 
- \sqrt{-g} \left( g^{00} \omega^2 + g^{11} q^2 + m^2 \right) X = \tilde{e} J \,,
\label{phi}
\ee
where the relevant Fourier-space components of the source are 
\be
J =  \sqrt{-\dot{x}^2 (r_0)}  \,  2\pi \delta( \omega - q v \cos \theta ) \, \delta (r - r_0)  \,,
\label{Jphi}
\ee
and as usual $\tilde{e}=e/\Omega_3$. We solve \eqn{phi} by expanding $X$ as in eq.~\eqn{X}. In the presence of the source, the equation obeyed by the $X_n$ mode is 
\be
\left[ \omega^2 - \omega_{n}^2 (q) \right]  X_n (\omega,q) = \tilde{e} J_n (\omega,q)\,,
\label{waveXwith}
\ee
where
\be
J_n (\omega, q) = \int dr J(\omega, q) \varphi_n (q,r) 
=  \sqrt{-\dot{x}^2 (r_0)}  \,  2\pi \delta( \omega - q v \cos \theta ) \varphi_n (q,r_0) \,.
\label{JnX}
\ee
With retarded boundary conditions, as appropriate for the reaction to the quark's passage, eq.~\eqn{waveXwith} yields 
\be
X_n(\omega,q) = 
\frac{\tilde{e} J_n (\omega, q)}{\left( \omega + i\epsilon \right)^2 - \omega_n^2(q)} \,.
\ee  
As in the case of vector modes, we see from eq.~\eqn{waveXwith} that the effective coupling between a scalar meson $X_n(\omega,q)$ and the quark is determined by the radial wave function of the meson, in this case 
\be
e_\mt{eff}(q,r_0)=e \,  \sqrt{-\dot{x}^2 (r_0)} \, \varphi_n (q,r_0) \,.
\label{eeffScalar}
\ee

We are now ready to compute the rate of energy deposition into scalar mesons. From 
eqs.~\eqn{formula} and \eqn{non} we have 
\be
\frac{dE_\mt{scalar}}{dt} = -e \int d^3 x dr d\Omega_3 \, \dot{X} J 
= - e   \sqrt{-\dot{x}^2 (r_0)} \, \dot{X} (t, \vec{v} t, r_0, \Omega_0) \,.
\label{Xloss}
\ee 
Following the steps of the vector meson case we find:
\bea
\frac{dE_\mt{scalar}}{dt} 
&=& \sum_n - e   \sqrt{-\dot{x}^2 (r_0)} 
\int \frac{d\omega d^3q}{(2\pi)^4} \left( -i\omega \right) 
X_n (\omega,q) \varphi_n(q,r_0)  \, e^{-i\omega t} e^{i t q\cdot v} \nn
&=& \sum_n - e  \left[ -\dot{x}^2 (r_0) \right] 
\int \frac{d^3q}{(2\pi)^3} \left( -iqv \cos \theta \right) 
\frac{\tilde{e}}{\left( q v\cos\theta + i\epsilon \right)^2 - \omega_n^2(q)} \,
\varphi_n^2(q,r_0) \nn
&=& \sum_n -\frac{e^2}{\Omega_3  v}  \left[ -\dot{x}^2 (r_0) \right]  
\int_0^\infty \frac{dq}{2\pi} \, q \, \varphi_n^2 (q,r_0) \int_{-1}^1 \frac{dz}{2\pi i} 
\frac{z}{(z+i\epsilon)^2 - z_n^2(q)} \nn 
&=& \sum_n \frac{e^2}{2 \Omega_3  v}  \left[ -\dot{x}^2 (r_0) \right]   
\int_0^\infty \frac{dq}{2\pi} \, q \, \varphi_n^2 (q,r_0) \, \Theta \left( 1 - \frac{v_n^2(q)}{v} \right) \,.
\label{scalarloss}
\eea
The result for the energy loss into the lowest-lying scalar and pseudoscalar modes is shown  in figs.~\ref{result-scalar} and \ref{result-pseudo}. The main difference with respect to the case of vector mesons is the fact that the constant-$r_0$ curves do not rise monotonically as $v$ increases, but instead they vanish when $v$ reaches the local speed of light at $r_0$. The reason for this is of course the factor in eq.~\eqn{sqrt}.  
\FIGURE{
\begin{tabular}{ccc}
\,\,\,\,\,\,\,\,
\includegraphics[width=0.42 \textwidth]{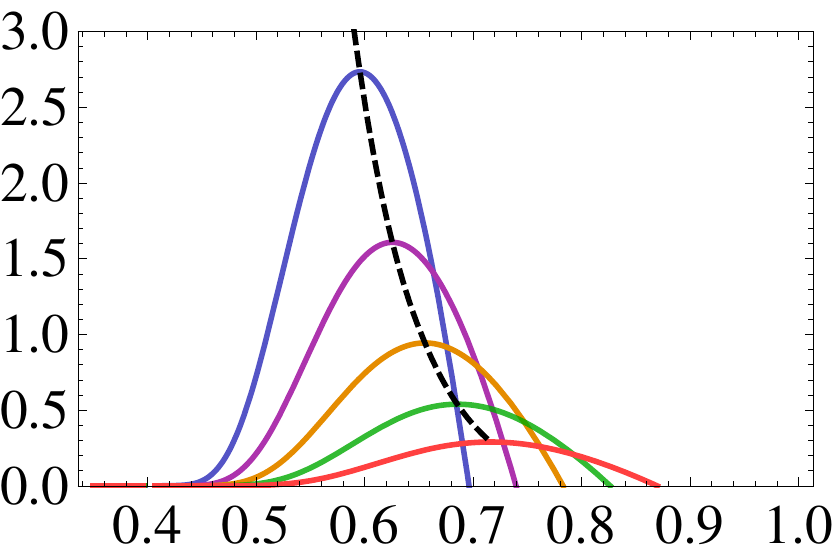} 
\put(-90,-10){$v$}
\put(-207,40){\rotatebox{90}{{$\frac{\nc}{\left( 2\pi T \right)^2} \frac{dE}{dt}$}}}
& \,\,\,\,\,\,\,\,\,\,\,\,\,&
\includegraphics[width=0.42 \textwidth]{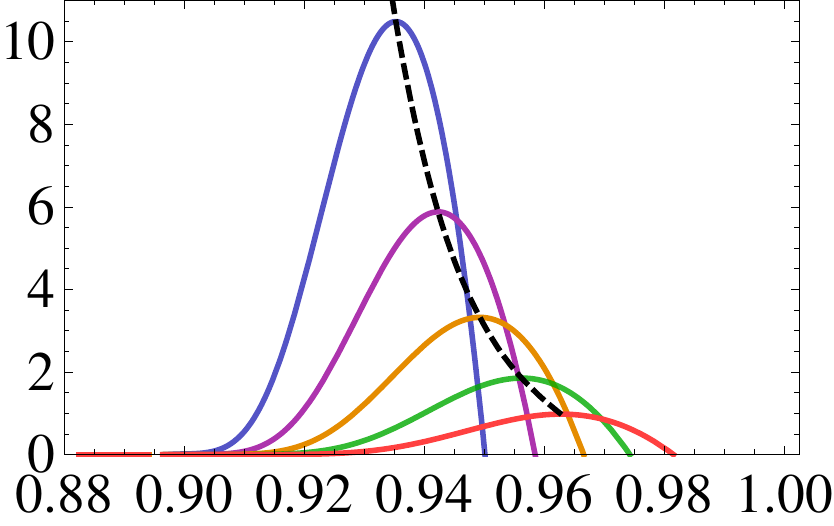} 
\put(-90,-10){$v$}
\put(-207,40){\rotatebox{90}{{$\frac{\nc}{\left( 2\pi T \right)^2} \frac{dE}{dt}$}}}
\end{tabular}
\caption{Energy loss into the scalar mode $\varphi_{(n=0)}^{(A=1)}$ for an embedding with $m=1.32, R_0=1.2$ (left) and $m=2.0, R_0=2.0$ (right). The continuous curves correspond (from top to bottom) to $r_0=0.86, 0.97, 1.10, 1.25, 1.45$ (left) and to 
$r_0=1.50, 1.69, 1.91, 2.18, 2.52$ (right). The dotted curve is defined by the maxima of the constant-$r_0$ curves.} 
\label{result-scalar}}
\FIGURE{
\begin{tabular}{ccc}
\,\,\,\,\,\,\,\,
\includegraphics[width=0.42 \textwidth]{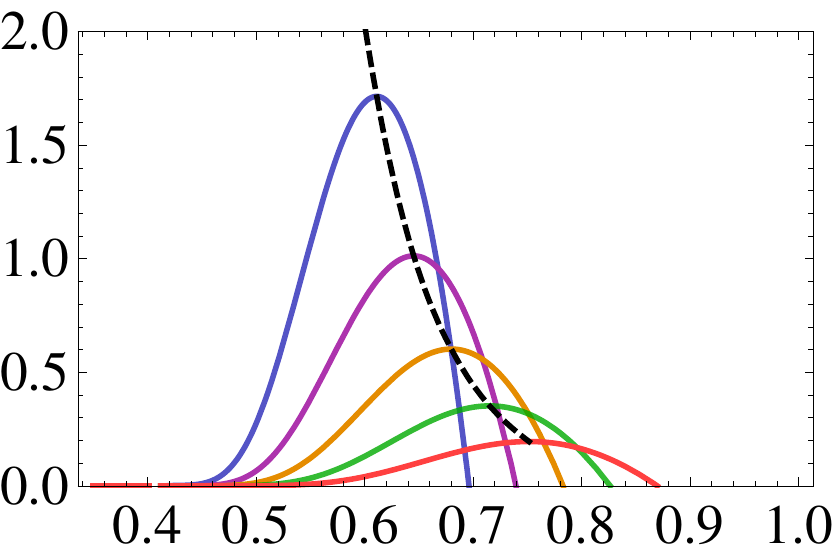} 
\put(-90,-10){$v$}
\put(-207,40){\rotatebox{90}{{$\frac{\nc}{\left( 2\pi T \right)^2} \frac{dE}{dt}$}}}
& \,\,\,\,\,\,\,\,\,\,\,\,\,&
\includegraphics[width=0.42 \textwidth]{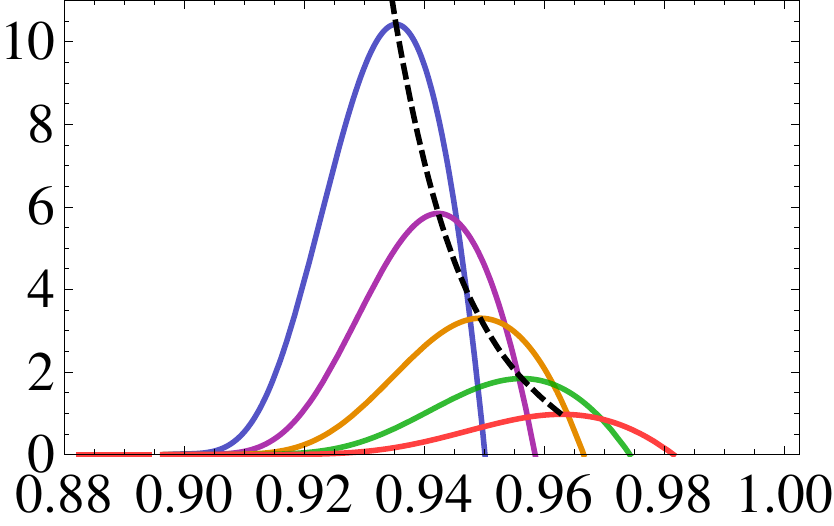} 
\put(-90,-10){$v$}
\put(-207,40){\rotatebox{90}{{$\frac{\nc}{\left( 2\pi T \right)^2} \frac{dE}{dt}$}}}
\end{tabular}
\caption{Energy loss into the pseudoscalar mode  $\varphi_{(n=0)}^{(A=2)}$ for an embedding with $m=1.32, R_0=1.2$ (left-hand side plots) and $m=2.0, R_0=2.0$ (right-hand side plots). The continuous curves correspond (from top to bottom) to $r_0=0.86, 0.97, 1.10, 1.25, 1.45$ (left) and to $r_0=1.50, 1.69, 1.91, 2.18, 2.52$ (right). The dotted curve is defined by the maxima of the constant-$r_0$ curves.} 
\label{result-pseudo}}

\section{Phenomenological implications for HIC experiments}
As is clear from our general discussion in section \ref{universal}, the mechanism of Cherenkov energy loss depends only on two qualitative properties encoded in the dispersion relations of fig.~\ref{dispersion}: the fact that heavy mesons remain bound in the gauge theory plasma, and the fact that their limiting velocity in the plasma is subluminal. Both properties can be motivated in QCD irrespectively of whether or not a string dual of QCD exists. The first property is suggested by the fact that sufficiently heavy mesons are smaller than the screening length in the plasma \cite{Matsui-Satz}, and is supported by calculations of both the static quark-antiquark potential \cite{potential} and of Minkowski-space spectral functions in lattice-regularized QCD \cite{correlators}.\footnote{In some models light mesons also remain bound above $T_c$ by Coulomb-like forces \cite{Shuryak:2004tx}.} The second property, which goes back to ref.~\cite{Chu-Matsui}, is suggested by the fact that moving mesons see a boosted, higher energy density that will melt them if they move sufficiently fast \cite{hotwind}.\footnote{An alternative possibility 
would be that meson states with $q$ above some upper bound cease to exist. In any event, our conclusions rely on meson states existing (with a sufficiently narrow width) only up to some moderate $q$ for which 
$q \gtrsim \omega(q)$.} 

Rigorously verifying these two properties in QCD is not presently feasible. For this reason it is reassuring that, as we explained in section \ref{universal}, they are both realized in all gauge theory plasmas with a gravity dual in the large-$N_c$, strong coupling limit. In this section we will assume that the two properties are also realized in the QGP and extract some phenomenological consequences that might be observable in heavy ion collisions.\footnote{Implications for photon production have been discussed in \cite{peak}, and for deep inelastic scattering in \cite{DIS}.} Since the heavier the meson the more perturbative its properties become, we expect that our conclusions are more likely to be applicable to the charmonium rather than to the bottomonium sector. 

An interesting feature of the energy loss by Cherenkov radiation is that, unlike other energy-loss mechanisms, it is largely independent of the details of the quark excited state, such as the precise features of the gluon cloud around the quark, etc. In the gravity description these details would be encoded in the precise profile of the entire string, but the Cherenkov emission only depends on the trajectory of the string endpoint. This leads to a dramatic simplification which, with the further approximation of rectilinear uniform motion, reduces the parameters controlling the energy loss to two simple ones: the string endpoint velocity $v$ and its radial position $r_0$. The former is just the velocity of the quark in the gauge theory, whereas the second roughly measures the size of the gluon cloud that dresses the quark \cite{size}. In order to obtain a ballpark estimate of the magnitude of the energy loss, we will assume that in a typical collision quarks are produced with order-one values of $r_0$. Under these circumstances the energy loss is of order unity in units of $(2\pi T)^2/N_c$, which for a temperature range of $T=200-400\,$ MeV and $N_c=3$ leads to $dE/dx \approx 2-8\,$ GeV/fm. This is is of the same order of magnitude as other mechanisms of energy loss in the plasma; for example, the BDMPS radiative energy loss $dE/dx=\alpha_s  C_F \hat q  L/2$ yields values of $dE/dx=7-40$ GeV/fm for $\hat q=1-5$ GeV$\,^2/$fm, $\alpha_s=0.3$ and $L\approx 6$ fm. Since our gravity calculation is strictly valid only in the infinite-quark energy limit (because of the linear trajectory approximation), we expect that our estimate is more likely to be applicable to highly energetic quarks at LHC rather than to those at RHIC. 

Even if in the QGP the magnitude of Cherenkov energy loss turns out to be subdominant with respect to other mechanisms, its velocity dependence and its geometric features may still make it identifiable. Indeed, Cherenkov energy loss would only occur for quarks moving at velocities $v > \vlim$, with $\vlim$ the limiting velocity of the corresponding meson in the plasma. The presence of such a velocity threshold is the defining characteristic of Cherenkov energy loss. The precise velocity at which the mechanism starts to operate may actually be higher than $\vlim$ in some cases, since the additional requirement that the energy of the quark be equal or larger than the in-medium mass of the meson must also be met. A related conclusion of our calculation in the 
${\cal N}=4$ model is that the energy loss decreases as the velocity of the quark approaches the speed of light. This is in fact a universal feature of all plasmas with a gravity dual. The reason is that the quark velocity can approach unity only in the limit 
$r_0 \rightarrow\infty$, in which the effective couplings \eqn{eeff}, \eqn{eeffLong} and \eqn{eeffScalar} between the quark and the mesons vanish because the meson radial wave functions are normalizable. In fact, in the case of scalar mesons the energy loss ceases completely at some subluminal velocity at which \eqn{sqrt} vanishes. If these properties also hold in QCD then Cherenkov energy loss may be identifiable because it only operates in a limited range of quark velocities.

Cherenkov mesons would  be radiated at a characteristic angle $\cos \theta_c=v_\mt{lim}/v$ with respect to the emitting quark, where $v$ is the velocity of the quark. Taking the gravity result as guidance, $\vlim$ could be as low as $\vlim=0.35$ at the meson dissociation temperature \cite{MMT2}, corresponding to an angle as large as 
$\theta_\mt{c} \approx 1.21$ rad. This would result in an excess of heavy mesons associated to high-energy quarks passing through the plasma. Our estimate of the energy loss suggests that the number of emitted $J/\psi$'s, for example, could range from one to three per fm. This emission pattern is similar to the emission of sound waves by an energetic parton \cite{mach} in that both effects lead to a non-trivial angular structure. One important difference, however, is that the radiated heavy mesons would not thermalize and hence would not be part of a hydrodynamic shock wave. As in the Mach cone case, the meson emission pattern could be reflected in azimuthal dihadron correlations triggered by a high-$p_\mt{T}$ hadron. Due to surface bias, the energetic parton in the triggered direction is hardly modified, while the one propagating in the opposite direction moves through a significant amount of medium, emitting heavy mesons. Thus, under the above assumptions, the dihadron distribution with an  associated $J/\psi$ would have a ring-like structure  peaked at an angle 
$\theta \approx \pi-\theta_\mt{c}$.

A final observation is that Cherenkov energy loss also has a non-trivial temperature dependence, since it requires that there are meson-like states in the plasma, and therefore it does not take place at temperatures above the meson dissociation temperature. Similarly, it is reasonable to assume that it does not occur at temperatures below $T_c$, since in this case we do not expect the meson dispersion relation to become spacelike.\footnote{This assumption is certainly correct for plasmas with a gravity dual, since the corresponding geometry does not include a black hole horizon if $T< T_c$.} Under these circumstances, the Cherenkov mechanism is only effective over a limited range of temperatures $T_c<T< T_\mt{diss}$ which, if $T_\mt{diss} \gtrsim 1.2 T_c$ as in \cite{Mocsy:2007yj}, is a narrow interval. As was pointed out in \cite{Liao:2008dk}, a mechanism of energy loss which is confined to a narrow range of temperatures in the vicinity of $T_c$ concentrates the emission of energetic probes to a very narrow layer on the collision geometry and is able to explain $v_2$-data at high $p_T$  at RHIC \cite{data}. Provided that the meson dissociation temperature $T_\mt{diss}$ is not much larger than $T_c$, the radiation of Cherenkov mesons is one such mechanism.

\section{Discussion}
\label{discussion}
\noindent
Cherenkov emission of mesons \cite{mesons} and gluons \cite{gluons} in QCD has been considered before. Although some of the underlying physics is similar, the mechanism we have discussed is different in several respects. First, it operates in the QGP, as opposed to in a hadronic medium as in \cite{mesons}, and the radiated particles are colourless mesons, as opposed to gluons as in \cite{gluons}. Second, the gauge/string duality provides a large class of completely explicit examples (although none of them includes QCD) in which this mechanism is realized and in which the energy loss can be calculated without further model assumptions. 

Figs.~\ref{result}, \ref{result-long}, \ref{result-scalar} and \ref{result-pseudo} show the rate of energy deposition into vector and scalar modes on the brane. All these figures share the property that the energy loss diverges as $1/r_0^6$ (as shown analytically in appendix   \ref{divergence}) in the limit $r_0 \rightarrow 0$. However, this mathematical divergence is removed by physical effects we have not taken into account. For example, for sufficiently large $q$ the radial profile of the mesons becomes of order the string length and stringy effects become important \cite{mit}. Also, mesons acquire widths $\Gamma \propto q^2$ at large $q$ \cite{width} and can no longer be treated as well defined quasiparticles. Finally,
the approximation of a constant-$v$, constant-$r_0$ trajectory ceases to be valid whenever the energy loss rate becomes large. 
\FIGURE{
\,\,\,\,\,
\includegraphics[width=0.85 \textwidth]{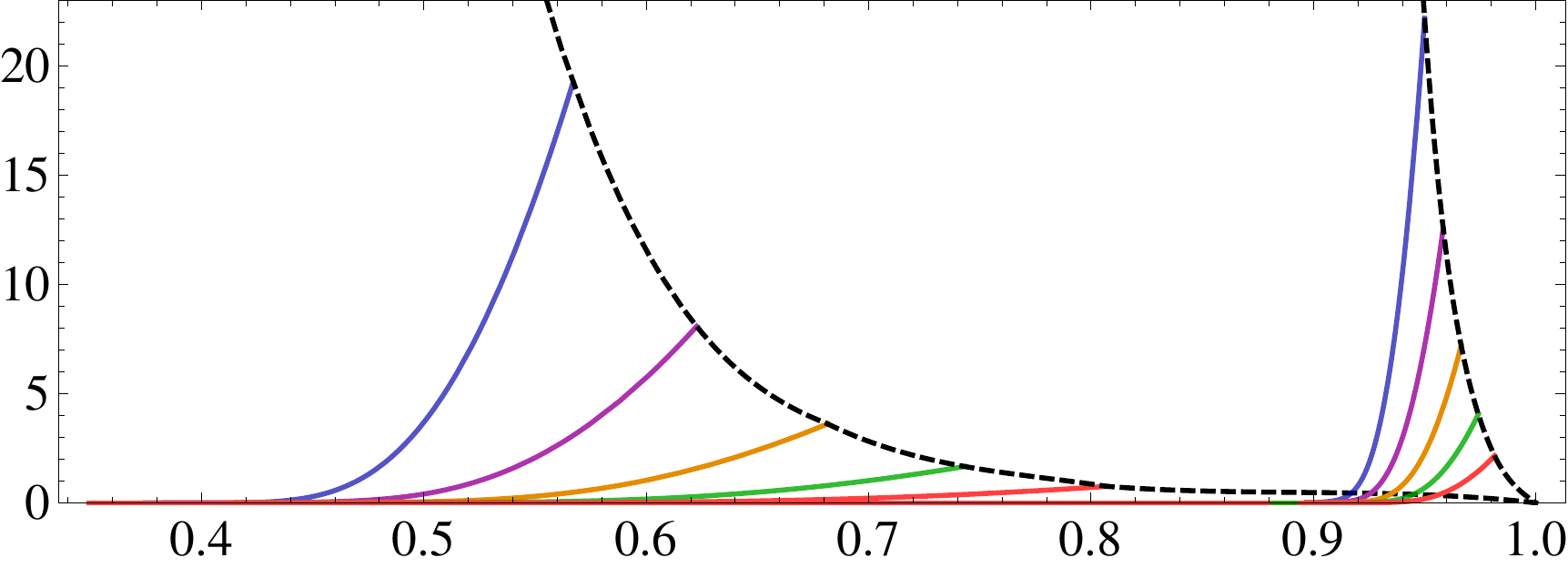} 
\put(-190,-10){$v$}
\put(-400,50){\rotatebox{90}{{$\frac{\nc}{\left( 2\pi T \right)^2} \frac{dE}{dt}$}}}
\caption{Energy loss into the transverse vector mode ${\cal A}_{(n=0)}$ for an embedding with $m=1.32, R_0=1.2$ (left-hand side curves) and $m=2.0, R_0=2.0$ (right-hand side curves). The continuous curves correspond (from top to bottom) to $r_0= 0.58, 0.70, 0.83, 0.98, 1.18$ (left-hand side) and to $r_0= 1.50, 1.69, 1.91, 2.18, 2.52$ (right-hand side). The dotted curve is defined by the endpoints of the constant-$r_0$ curves.}
\label{result-combined1}}
\FIGURE{
\,\,\,\,\,
\includegraphics[width=0.85 \textwidth]{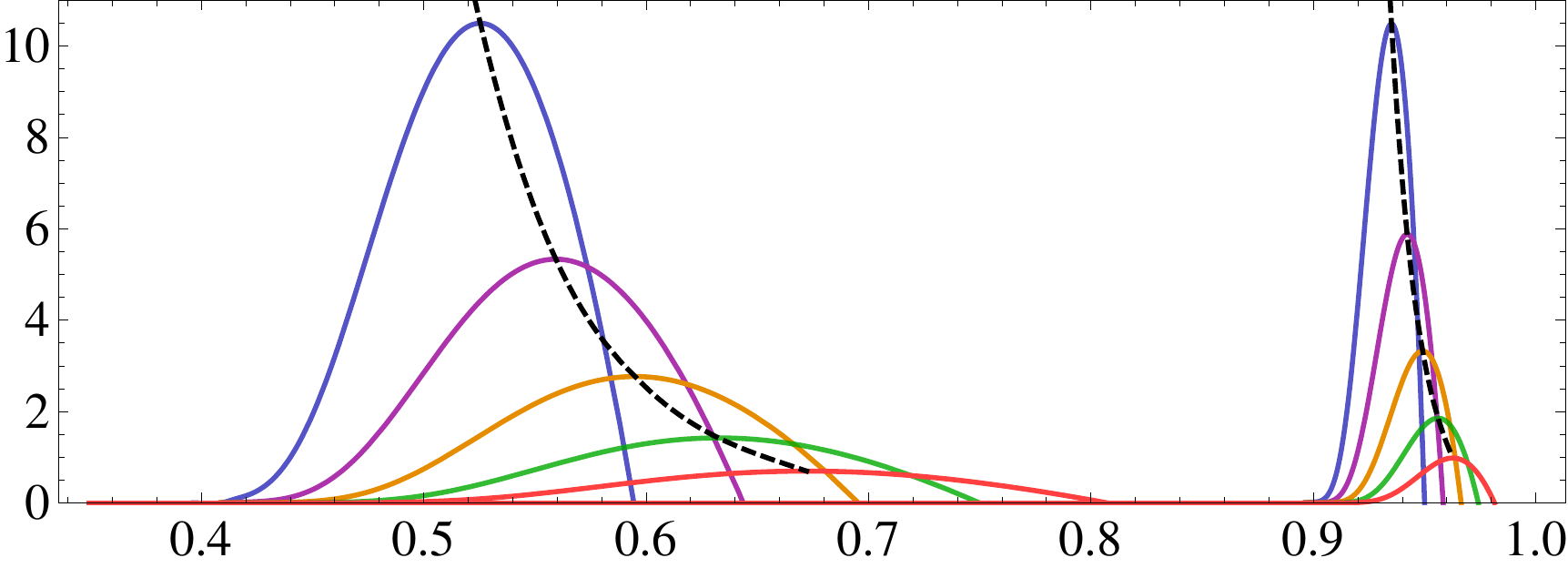} 
\put(-190,-10){$v$}
\put(-400,50){\rotatebox{90}{{$\frac{\nc}{\left( 2\pi T \right)^2} \frac{dE}{dt}$}}}
\\
\caption{Energy loss into the scalar mode $\varphi_{(n=0)}^{(A=1)}$ for an embedding with $m=1.32, R_0=1.2$ (left-hand side curves) and $m=2.0, R_0=2.0$ (right-hand side curves). The continuous curves correspond (from top to bottom) to $r_0= 0.64, 0.74, 0.86, 1.00, 1.18$ (left-hand side) and to $r_0=1.50, 1.69, 1.91, 2.18, 2.52$ (right-hand side). The dotted curve is defined by the maxima of the constant-$r_0$ curves.}
\label{result-combined2}}

Figs.~\ref{result}, \ref{result-long}, \ref{result-scalar} and \ref{result-pseudo} also illustrate the simple dependence of the energy loss on the ratio $m \propto M_q/T$. Increasing $m$ decreases the redshift at the bottom of the branes, and therefore increases the limiting velocity of mesons, $\vlim$, at which quark energy loss via Cherenkov emission starts to operate. This means that the energy loss becomes concentrated on a narrower range of velocities, closer to unity, as $m$ is increased, but the structure of the curves is roughly the same up to a rescaling. This can be seen in the figures above by comparing the energy loss for $m=1.2$ (left-hand side) and $m=2.0$ (right-hand side). The concentration of energy loss on a narrower velocity interval is also illustrated in figs.~\ref{result-combined1} and \ref{result-combined2}, where the result for both values of $m$ is shown simultaneously on the same plot (for the transverse vector and scalar modes). 

In this paper we have concentrated on the case $\nf=1$, i.e.~we have assumed the presence of a single heavy flavour. Consider now a theory with multiple heavy quarks, such as QCD with $c$ and $b$ quarks, for example. In the string description this corresponds to a situation with $\nf >1$ D-branes. If all quarks have identical masses (and R-symmetry quantum numbers) then the D-branes are all coincident and their worldvolume theory is described by a non-Abelian $U(\nf)$ theory, corresponding to the fact that mesons $m^{f\tilde{f}}$ come in multiplets that transform in the adjoint representation of $U(\nf)$. A quark with flavour $f$ then may emit any of the $\nf$ mesons with flavour 
$f\tilde{f}$, with $\tilde{f}=1, \ldots, \nf$. Under these circumstances the energy loss is enhanced by a power of $\nf$.\footnote{This corrects the corresponding statement in 
ref.~\cite{prl}.} Note that if the resulting meson has $\tilde{f} \neq f$ then the quark must change flavour $f \rightarrow \tilde{f}$ in the emission process, as shown in fig.~\ref{emission}(left). A process in which the quark does not change flavour is also possible, as shown in fig.~\ref{emission}(right), but this requires the emission of at least two mesons and is therefore further suppressed at large 
$\nc$.
\FIGURE{
\includegraphics[width=0.82 \textwidth]{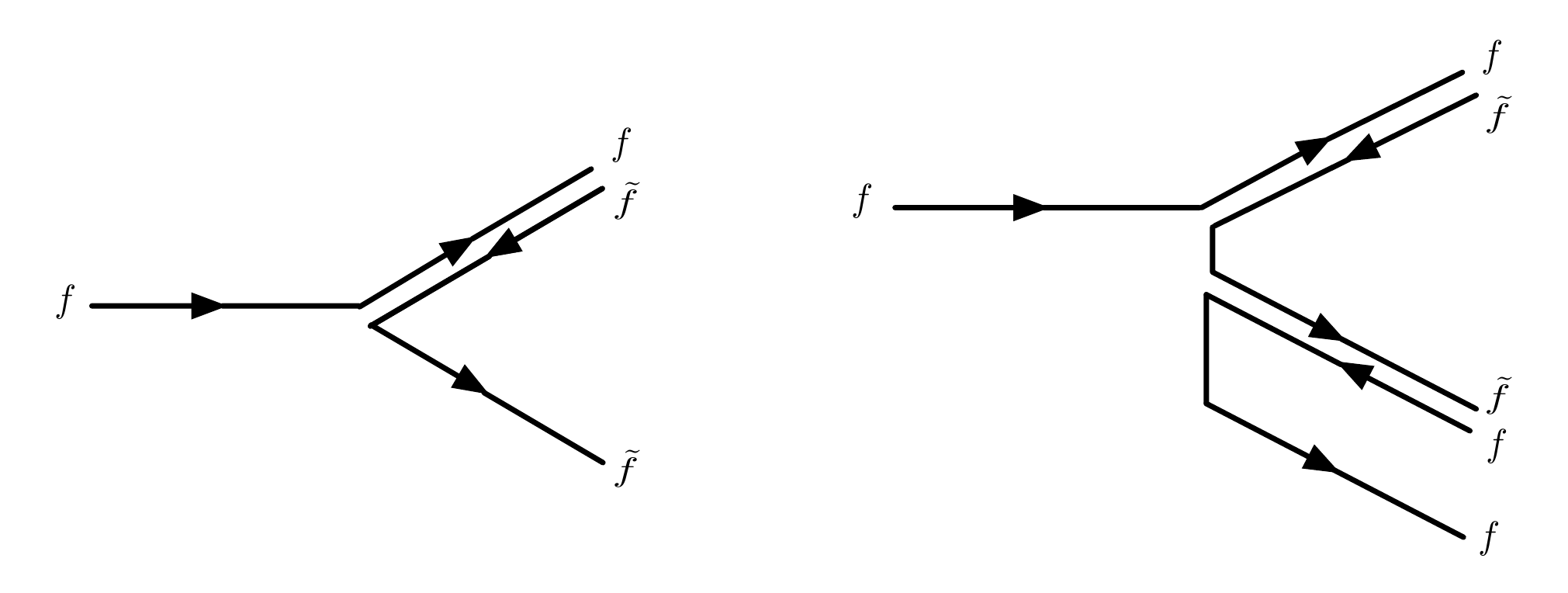} 
\caption{Meson emissions by a quark in a theory with multiple flavours.} 
\label{emission}}

Consider now the opposite situation, more analogous to QCD, in which the heavy quarks have different masses, so that the D-branes in the string description do not overlap. This is depicted in figs.~\ref{multiple1}, \ref{multiple2} and \ref{multiple3}, which show the string description of the emission processes of figs.~\ref{emission}(left) and \ref{emission}(right), respectively. Figs.~\ref{multiple1} and \ref{multiple2} correspond to emissions with $\tilde{f} \neq f$, whereas fig.~\ref{multiple3} describes the emission of an $ff$-meson. In this geometric picture the necessity of a two-meson emission in the case $\tilde{f} \neq f$ in order to preserve the quark flavour is due to the fact that the string must break twice in order to stay attached to the same brane -- see fig.~\ref{multiple2}. Since string breaking is suppressed at large $\nc$, this process is subleading with respect to one-meson emission. In any case, since an open string must always have its endpoints attached to a brane, the emission of one or multiple mesons with $\tilde{f} \neq f$ by a quark of flavour $f$ requires a tunneling process in which the string fluctuates and touches the $\tilde{f}$-brane, as shown in figs.~\ref{multiple1} and \ref{multiple2}. The amplitude for this process can be studied semiclassically provided the distance between the two branes  is sufficiently large compared to $\ell_s$, but it is far from straightforward to calculate \cite{exp}. In addition, in the present context the calculation would require a precise specification of the string profile. On general grounds, however, one may expect the amplitude to be exponentially suppressed, since it requires a large string fluctuation that is classically forbidden. Note that the same exponential suppression applies to the emission of a large $ff$-meson by an $f$-quark, as shown in fig.~\ref{multiple3}. This is the reason why we neglected this process in our calculation of energy loss. More precisely, our calculation can be seen as accounting for this process in the limit in which the size of the emitted meson is so small that it requires quantization of the resulting string. In this limit there is no exponential suppression, and the emitted string must be described as a field propagating on the brane.
\FIGURE{
\includegraphics[scale=.50]{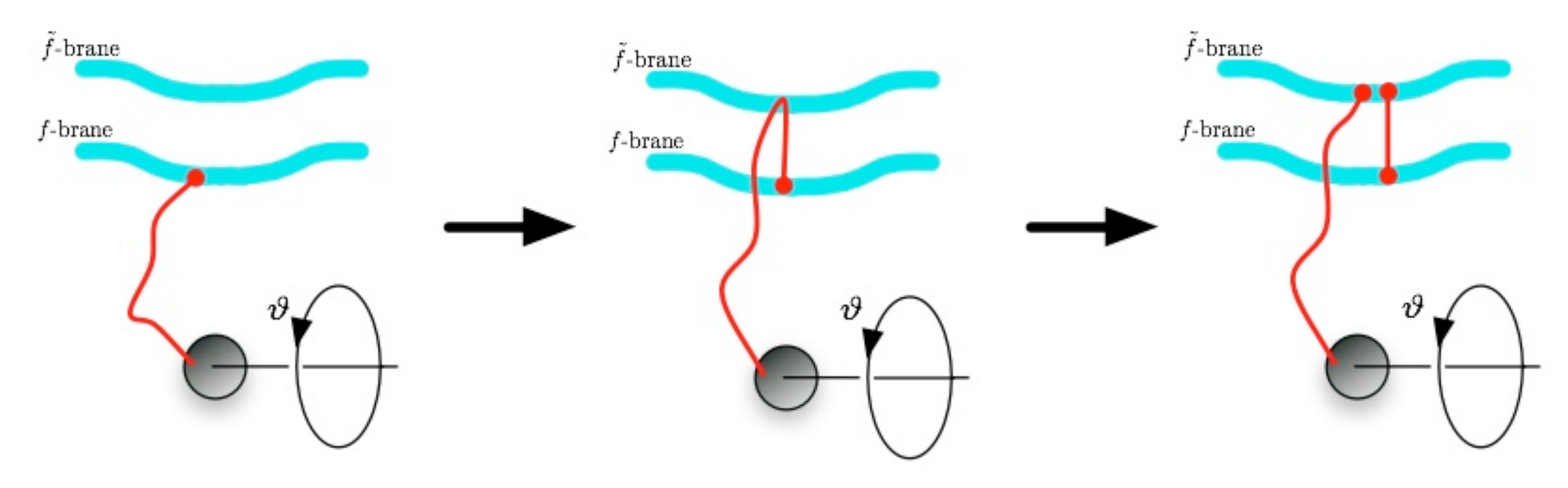}
\caption{String description of the emission of an $f\tilde{f}$-meson with $\tilde{f} \neq f$. The quark changes flavour in the process. } 
\label{multiple1}}
\FIGURE{
\includegraphics[scale=.50]{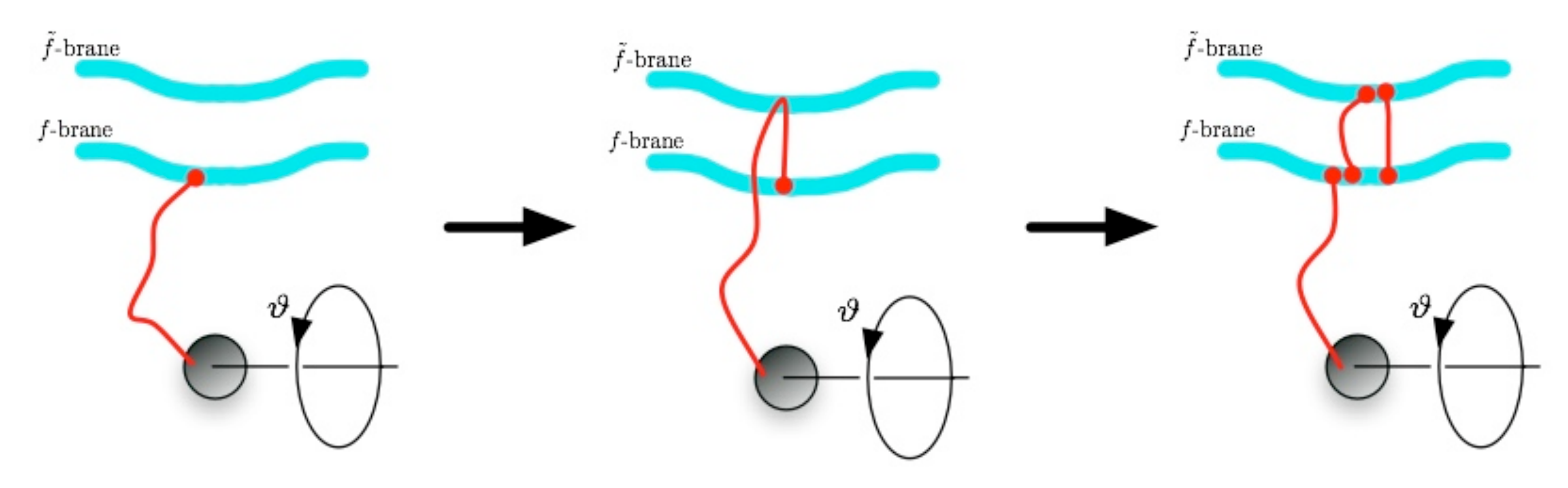}
\caption{String description of the emission of two mesons with $\tilde{f} \neq f$. The quark does not change flavour in the process.} 
\label{multiple2}}
\FIGURE{
\includegraphics[scale=.50]{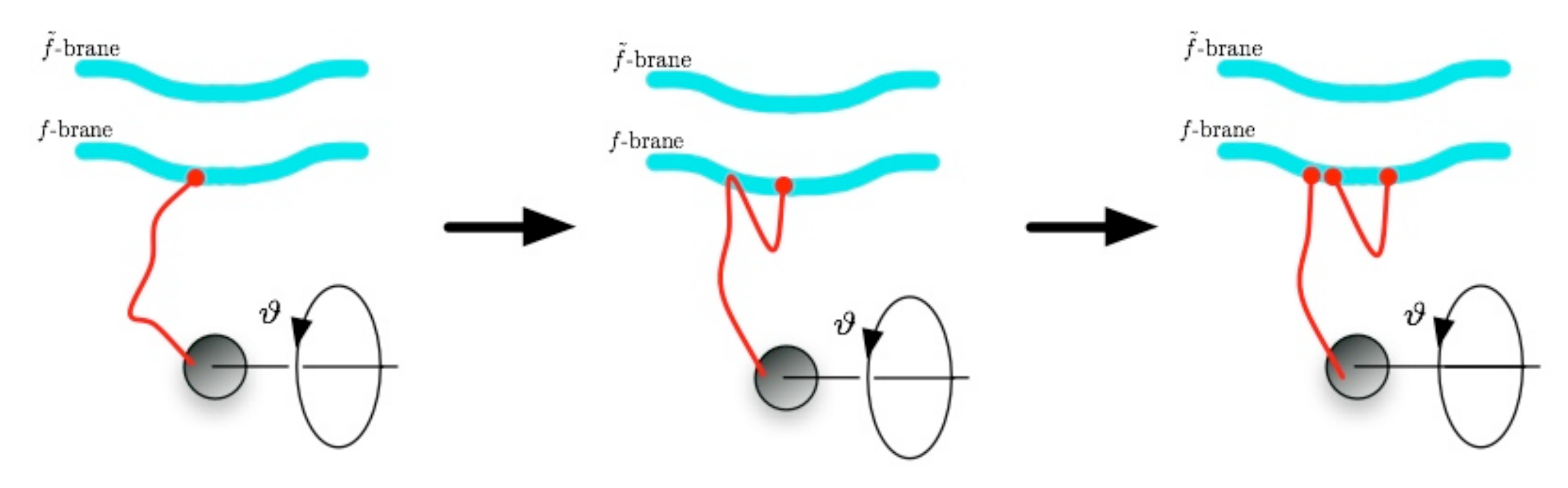}
\caption{String description of the emission of an $ff$-meson by an $f$-quark. The quark does not change flavour in the process.} 
\label{multiple3}}

We close with a comment on a possible extension of our work. In this paper we have focused on the energy loss of quarks attached to branes that sit outside the horizon in a Minkowski embedding. It would be interesting to study the energy loss in the case of black hole embeddings, which describe light quarks. In this case no stable quark-antiquark bound states exist in the plasma, which in the string description corresponds to the fact that excitations on the brane are characterized by quasinormal modes with complex frequencies. It would be interesting to explore whether the emission of quasinormal modes could also lead to a significant energy loss.

\acknowledgments 
We thank Mariano Chernicoff, David d'Enterria, Roberto Emparan, Bartomeu Fiol, Alberto Guijosa, Cristina Manuel, Angel Paredes, Robert Myers, Leonardo Pati\~{n}o, Amit Sever and Paul Townsend for discussions. We are supported by a Marie Curie
Intra-European Fellowship 
PIEF-GA-2008-220207 (JCS) and by 2009-SGR-168, MEC FPA 2007-66665-C02 and CPAN CSD2007-00042 Consolider-Ingenio 2010 (DF, DM).

\appendix
\section{Boundary conditions at the string endpoint}
\label{derivation}
The action for the string may be written as 
\be
S = -T_\mt{string} \int d\tau d\sigma \sqrt{-g} \, \frac{1}{2} \,g^{\alpha\beta} 
\partial_\alpha X^M \partial_\beta X^N G_{MN} \,.
\ee
In this form of the action $g$ is an {\it independent} worldsheet metric and 
$X^M(\tau,\sigma)$ specify the embedding in spacetime of the string worldsheet. Since $g$ appears undifferentiated, it can be eliminated from the action through its equation of motion. This equation implies that 
\be
g_{\alpha\beta} =  \partial_\alpha X^M \partial_\beta X^N G_{MN} \,,
\ee
i.e.~that $g$ is the induced metric on the worldsheet. Substituting this into the action one obtains the familiar Nambu-Goto action
\be
S = -\frac{T_\mt{string}}{2} \int d\tau d\sigma \sqrt{-g} \,.
\ee

An alternative way to proceed, which is more convenient to elucidate the boundary conditions at the string endpoint, is to choose the so-called conformal gauge. This means that one uses the reparametrization invariance of the string action to ensure that the worldsheet metric is conformally flat, i.e~that 
$g_{\alpha\beta} = \Omega^2(\tau,\sigma)\eta_{\alpha\beta}$. (In addition, Weyl invariance may be used to ensure that $\Omega=1$.) In this gauge the action becomes
\be
S = -T_\mt{string} \int d\tau d\sigma \frac{1}{2} \,\eta^{\alpha\beta} 
\partial_\alpha X^M \partial_\beta X^N G_{MN} \,.
\ee
Variation of this action with respect to the embedding coordinates yields a bulk term proportional to the equation of motion, $\eta^{\alpha\beta} \partial_\alpha \partial_\beta X^M =0$, plus the boundary term 
\be
-T_\mt{string} \int d\tau \left[  {X^{M}}^\prime \delta X^N G_{MN} \right]_\mt{bdry} \,,
\ee
which is integrated over the string boundary. The equation of motion must be supplemented by the constraints associated to the gauge fixing of $g$, which take the form 
\bea
\dot{X} \cdot {X^{\prime}} &\equiv& G_{MN} \dot{X}^M {X^{N}}^\prime = 0 \,, \nn
\dot{X}^2 + {X^{\prime}}^2 &\equiv& 
G_{MN} \left( \dot{X}^M \dot{X}^N + {X^{M}}^\prime {X^{N}}^\prime \right) = 0 \,.
\eea
The boundary conditions follow from the requirement that the boundary term vanish. This may be achieved by imposing either a Neumann boundary condition, 
${X^{M}}^\prime |_\mt{bdry} = 0$, or a Dirichlet boundary condition, 
$\delta X^{M} |_\mt{bdry}=0$. If all coordinates satisfy Neumann boundary conditions, then the second constraint immediately implies that $\dot{X}^2 |_\mt{bdry}=0$, namely the familiar condition that the endpoint moves at the speed of light. Suppose however that the string is attached to a Dp-brane, and let the first $p+1$ coordinates $X^a$ be coordinates along the brane directions, and $X^A$ be coordinates orthogonal to the brane. Then by definition we must choose Neumann boundary conditions for $X^a$ and Dirichlet boundary conditions for $X^A$: 
\be
{X^{a}}^\prime |_\mt{bdry} = 0 \sac \dot{X}^A  |_\mt{bdry} = 0 \,.
\ee
The Neumann boundary condition on $X^a$ implies that the string ends orthogonally on the brane, since the vector tangent to the string at its endpoint, 
${X^{M}}^\prime |_\mt{bdry} = 0$, has no components along the brane. Substituting both boundary conditions on the second constraint equation one finds that
\be
\left. {{\dot{X}}^a} \right.^2 |_\mt{bdry}= - \left. {X^{A}}^\prime \right.^2 |_\mt{bdry} 
\leq 0 \,,
\ee
which means that the endpoint moves along the brane at a speed lower than or equal to the local speed of light.

\section{Energy loss formula}
\label{loss}
A charge can be defined for each of the isometries of the brane's worldvolume metric \eqn{metric}. For concreteness, in this section we will focus on the four-momentum associated to translations in the gauge theory, which is generated by the set of Killing vector fields $k_\mu = \partial_\mu$. 

Let $\Sigma$ be a spacelike 7-surface in the brane's worldvolume, which we take to be a $t=\mbox{const.}$ surface, and $n = \partial_t / \sqrt{-g_{00}}$ the future-pointing unit normal to $\Sigma$. The momenta are then given by 
\be
P_\mu = \int_\Sigma d^7 x \, \sqrt{g_\mt{sp}} \, n^a \,T_{ab} \,k^b_\mu = 
\int_\Sigma d^7 x \, \sqrt{g_\mt{sp}} \, n^a \,T_{a\mu} \,,
\label{pmu}
\ee
where $g_\mt{sp}$ is the spatial metric on $\Sigma$. Since the time-space off-diagonal components of $g$ vanish we have that 
\be
g^{00} = 1/g_{00} \sac \sqrt{-g} = \sqrt{-g_{00}} \sqrt{g_\mt{sp}} \,.
\ee
Using these relations $P_\mu$ may be rewritten as 
\be
P_\mu =  \int_\Sigma d^7 x \, \sqrt{g_\mt{sp}} \, \frac{1}{\sqrt{-g_{00}}} \, T_{0\mu} =
- \int_\Sigma d^7 x \, \sqrt{-g} \, T^0_{\,\,\,\,\mu} \,.
\ee
For $\mu=0$, these formulas give the energy on the brane:
\be
E=P_0=  \int_\Sigma d^7 x \, \sqrt{g_\mt{sp}} \, \frac{1}{\sqrt{-g_{00}}} \, T_{00} = 
- \int_\Sigma d^7 x \, \sqrt{-g} \, T^0_{\,\,\,\,0} \geq 0 \,.
\ee
Note that this is non-negative because $T_{00} \geq 0$.\footnote{For example, in flat space  eq.~\eqn{Tvector} gives $T_{00} = E^2/2 + B^2/2  \geq 0$, where $E_i = F_{i0}$, $E^i = F_{\,\,0}^{i}$, $E^2 = E_i E^i$, and $2B^2 = F_{ij} F^{ij}$.}

Consider now the brane's worldvolume $V$ as shown in fig.~\ref{volume}. 
$\Sigma_1$ and $\Sigma_2$ are spacelike hypersurfaces at times $t_1$ and $t_2$, respectively, and $\Sigma_\infty$ is a timelike hypersurface at spatial infinity. Applying Stoke's theorem we then have
\be
\int_V \sqrt{-g} \, \nabla^a T_{a\mu} = 
\int_{\Sigma_2} \sqrt{g_\mt{sp}}\, n^a_{(1)} T_{a\mu} +
\int_{\Sigma_1} \sqrt{g_\mt{sp}}\, n^a_{(2)} T_{a\mu} + 
\int_{\Sigma_\infty} \sqrt{g_\mt{sp}}\, n^a_{(\infty)} T_{a\mu} \,.
\label{stoke}
\ee
\FIGURE{
\,\,\,\,
\includegraphics[scale=.30]{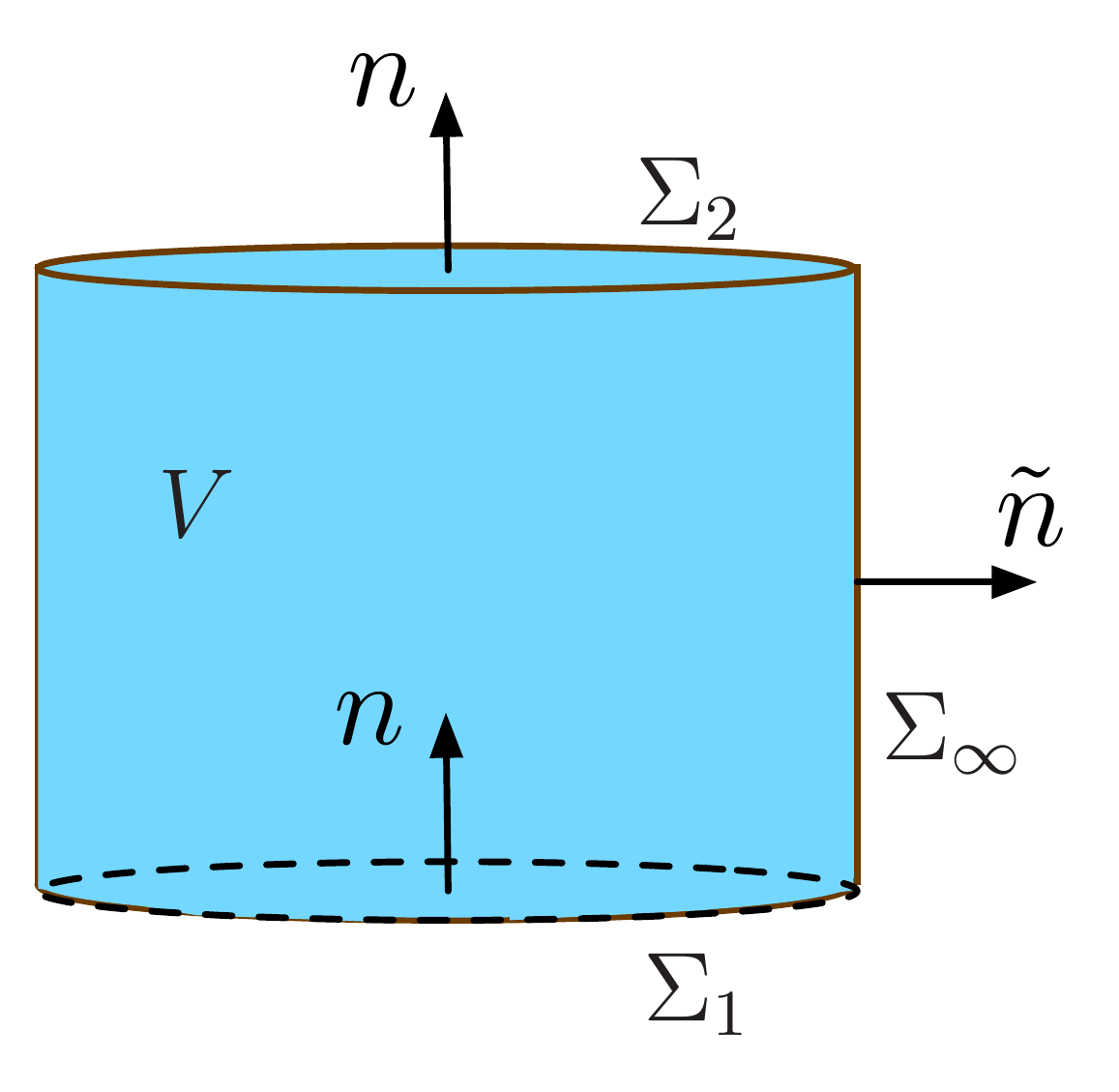}
\caption{Brane's worldvolume, to which Stoke's theorem is applied.} 
\label{volume}}
\noindent
The unit normals must be taken inward-pointing if they are time-like, and outward-pointing if they are space-like. Therefore we have $n^a_{(1)} = - n^a_{(2)} = n$ and 
$n^a_{(\infty)} = \tilde{n}$, so eq.~\eqn{stoke} with $\mu=0$ yields
\be
E_2 - E_1 + F_\infty = - \int_V \sqrt{-g} \, \nabla^a T_{a0} \,,
\label{source}
\ee
where $E_{1,2}$ is the energy contained in $\Sigma_{1,2}$ and 
$F_\infty=-\int_{\Sigma_\infty} \sqrt{g_\mt{sp}}\, \tilde{n}^a T_{a0}$ is the energy flux that has escaped through $\Sigma_\infty$ between $t_1$ and $t_2$.\footnote{The minus sign in the definition of the flux comes from the fact that the energy current in the $a$-direction is given by $-T_{a0}$, as can be seen from the continuity equation. For example, in flat space   in the absence of sources the continuity equation $\partial^a T_{a0} = 0$ yields 
$\partial_0 T_{00} - \partial_i T_{i0} = 0$.} The left-hand side of eq.~\eqn{source} is the total energy deposited on the brane by the source. Since $\int_V = \int dt d^7 x$, if $t_1$ and $t_2$ are inÞnitesimally close we obtain 
\be
\frac{dE}{dt} = - \int d^7 x \sqrt{-g} \, \nabla^a T_{a0} \,.
\ee

\section{High-momentum radial wave functions}
\label{hmwf} 
As shown (on the left-hand side of) figs.~\ref{transprofiles}, \ref{longprofiles}, and \ref{scalarprofiles}, the high-momentum radial profiles of the different modes are concentrated near the tip of the brane ($r=0$). In this region of high $q$ and small $r$ it is possible to find analytic expressions for the radial profiles \cite{mit}. In this appendix we shortly review this computation and extend it  to vector modes.

Following \cite{mit} we introduce the coordinate $z$ which fulfills
\be
\label{zorgen}
\frac{dz}{dr} =\sqrt{\frac{g_{rr}}{-g_{00}}}=\sqrt{\frac{2 \tildef  (1+\dot R^2)}{\rho^4 f^2}} \,.
\ee
In terms of this new coordinate, the different modes $\Psi^\alpha=\{ \phi_1,\, \phi_2 ,\,\mathcal{A},\, \Phi\}$ satisfy a differential equation of the generic form
\be
\label{eqngen}
\prt_z \left[ a^\alpha(z) \, \prt_z\Psi^\alpha \right] + 
a^\alpha(z) \left[ \omega^2- \frac{f^2}{\tildef ^2} q^2 \right] \Psi^\alpha -
b^\alpha (z) \, (m^\alpha)^2 \Psi^\alpha=0 \,,
\ee
where $m^\alpha=\{m_{11},\, m_{22},0\,,0\,\}$ and the different coefficient functions are given by
\bea
a^1=a^2 &=& -\sqrt{-g} \, g^{00} \,, \nn 
a^{3} &=& -\sqrt{-g}\, g^{00} g^{11} \,, \nn
a^{4} &=& - g_{11} g_{00}/\sqrt{-g} \,,\nn
b^1=b^2 &=& \sqrt{-g}\,.
\eea
Via the simple rescaling $\psi^\alpha=\sqrt{a^\alpha} \, \Psi^\alpha$, the set of eqs.~\equ{eqngen} can be written in  the Schr\"odinger form 
\be
-\prt^2_z \, \psi^\alpha + V^\alpha \, \psi^\alpha=\omega^2 \psi^\alpha \, ,
\label{potgeng}
\ee
where the potential is given by
\be
\label{potgen}
V^{\alpha}(z,q)=q^2 \, \frac{f^2}{\tilde f^2} + 
\frac{\prt^2_z\left(\sqrt{a^\alpha}\right)}{\sqrt{a^\alpha}}
+\frac{b^\alpha}{a^\alpha} (m^\alpha)^2\,.
\ee
This potential is a complicated function of $z$ which is only known numerically because it depends non-trivially on the brane embedding. However, since the radial profiles are concentrated near the tip of the brane, we focus on the small-$z$ expansion of the potential \equ{potgen}. In the limit $z\rightarrow0$, eq.~\equ{zorgen} reduces to\footnote{To derive this expression we have used the small-$r$ expansion of the embedding, $R(r)=R_0+r^2/\left(R_0(R^8_0-1)\right)$.}
\be
\label{zofrlin}
z=\frac{\sqrt{2 (R_0^4+1)}}{R_0^4-1} \, r \,,
\ee
and the small-$z$ potential is given by
\be
\label{eqho}
V^\alpha(z,q)=\left(\frac{3 }{4} + \ell^\alpha(\ell^\alpha+2) \right)\frac{1}{z^2} + 
v^2_\mt{lim} q^2 + \frac{1}{4} z^2 q^2 \Omega^2  \, ,
\ee
where $\ell^\alpha=\{0,0,0,1\}$, $v_\mt{lim}$ is the meson limiting velocity \eqn{vlim}, and
\be
\Omega^2=\frac{16 R_0^2 (R_0^4-1)^2(1+R_0^8)}{(1+R_0^4)^5} \, .
\ee
As claimed, this potential has a minimum at $z\propto1/\sqrt{q}$ which means that, at least for the lowest modes, the wave functions are concentrated at small $z$.  Different meson excitations correspond to  different states of the four-dimensional harmonic oscillator \equ{eqho}. The eigenfunctions and eigenvalues are given by
\bea
\label{drhq}
\omega^2_n&=&v_\mt{lim}^2 q^2 + q  \Omega (2n+ 2+ \ell^\alpha)\,,\\
\psi^\alpha_n&=&
\mathcal{N}^\alpha z^{\frac{3}{2}+\ell^\alpha} L^{\left(\ell^\alpha+1\right)}_n\left(\frac{1}{2}\Omega q z^2\right)
\exp\left(-\frac{1}{4} \Omega q z^2\right)\,,
\eea
where  $L_n^{\ell+1}$ is the generalized Laguerre polynomial and $\mathcal{N}^\alpha$ are normalization constants determined by the requirements
\be
\int dz \, \psi^\alpha_n(z) \, \psi^\alpha_m(z)=
\delta_{mn} \,.
\ee
These normalization conditions coincide with those in eqs.~\eqn{ortho}, \eqn{phiortho} and \eqn{varphiortho}.

For future use, we provide explicit expressions for the lowest excitations, which correspond to $n=0$. The normalization constants are given by
\be
\mathcal{N^\alpha}=\frac{\left(q \Omega\right)^{1+\frac{\ell^\alpha}{2}}}{\sqrt{2}^{1+\ell^\alpha} \sqrt{\left(1+ \ell^\alpha\right)!}}\,,
\ee
and the radial wave functions take the form
\be
\label{hqfs}
\Psi^\alpha_0=\beta^\alpha \mathcal{N}^\alpha z^{4 \ell^\alpha} \exp\left(-\frac{1}{4} \Omega q z^2\right) \, ,
\ee
where $\beta^\alpha = \lim_{r \rightarrow 0} \sqrt{a^\alpha}$:
\bea
\beta^1= \beta^2 &=&\frac{2^{3/2} R_0^3}{\left(R_0^4-1\right)^{3/2}} \, ,
\\
\beta^3&=&\frac{2 R_0^2\left(1+R_0^4\right)^{1/2}}{\left(R_0^4-1\right)^{3/2}} \, ,
\\
\beta^4&=&\frac{\left(R_0^4-1\right)^{3/2}}{2 R_0^2 \left(1+R^4_0\right)^{1/2}} \, .
\eea

\section{Energy loss at small $r_0$}
\label{divergence}
As the quark position approaches the tip of the branes, $r_0 \rightarrow 0^+$, the maximum velocity of the quark approaches the meson limiting velocity $v_\mt{lim}$ from above. As a consequence, the quark and meson dispersion relations cross at $q_0\gg T$. In fig.~\ref{dispersion} this means that the dotted vertical lines move to the right. For a fixed $r_0$ the smallest value of the crossing point, $q^\mt{min}_0$, is attained at the maximal velocity of the quark, $v_\mt{max}$. Using the high-momentum dispersion relation \equ{drhq}, the crossing momentum at an arbitrary quark velocity $v$ is determined by the condition
\be
v_\mt{lim} q_0 + \left(1+n+ \frac{
\ell^\alpha}{2}\right)\frac{\Omega}{v_\mt{lim}}= v q_0  \,,
\ee
which leads to
\be
q_0=\left(1+n+\frac{\ell^\alpha}{2}\right) \frac{\Omega}{v_\mt{lim}} \, \frac{1}{v-v_\mt{lim}}  \,.
\label{q0}
\ee
The maximal velocity for a quark at $r_0$ is 
\be
\label{vmax}
v_\mt{max} (z_0)=\left. \sqrt{-\frac{g_{00}}{g_{ii}}} \right|_{r_0} \approx  
v_\mt{lim} + z_0^2 \, \frac{\Omega^2}{8 v_\mt{lim}}\,,
\ee
where we have expanded to leading order in $r_0$ and used the definition \equ{zofrlin} of $z$. Substituting this value of $v$ in eq.~\eqn{q0} we find that the minimum crossing point is
\be
\label{qogen}
q^\mt{min}_0=\left(1+n+\frac{\ell^\alpha}{2}\right)\frac{8}{\Omega  z_0^2}\,.
\ee
Thus, the energy loss can be reliably computed with the approximate solutions of appendix \ref{hmwf}. Since the energy-loss formulas are different for each mode, we will address them separately. Furthermore, we will focus on the lowest state of each mode.

\subsection{Scalar mesons}
Since in the high-momentum, small-$r$ region both scalar modes have the same radial profile, the energy loss into scalar mesons in this limit will also be the same. Expanding the energy loss formula \eqn{scalarloss} and using the radial profiles \equ{hqfs} we obtain
\be
\frac{dE_\mt{scalar}}{dt} = \frac{e^2 }{2\Omega_3}\, 
\left[ \frac{v^2_\mt{max}(z_0) -v^2}{v} \right] 
\left[ \frac{1+R_0^4}{2 R_0^2} \right]
\left(\beta^1\right)^2 \frac{\Omega^2}{2} 
\int^\infty_{q_0^\mt{min}} \frac{dq}{2\pi} \, q^3 \exp\left(-\frac{1}{2} \Omega q z_0^2\right) \, .
\ee
After integration, this yields
\bea
\frac{dE_\mt{scalar}}{dt}&=& \frac{e^2 }{4\pi\Omega_3}
\frac{\left(1+R^4_0\right)^2 \left(\beta^1\right)^2}{2 R_0^2} 
\left[ \frac{v^2_\mt{max}(z_0) -v^2}{v} \right] \times \nonumber \\
&& \quad
\left[ \frac{48+24\Omega q_0^\mt{min} z^2_0 + 6 \Omega^2 (q_0^\mt{min})^2 z_0^4 +
\Omega^3 (q_0^\mt{min})^3 z_0^6 }
{\Omega^2 z_0^8} \right]
\exp\left(-\frac{1}{2} \Omega q_0^\mt{min} z_0^2\right) \,.
\eea
As expected on general grounds, and in agreement with figs.~\ref{result-scalar} and \ref{result-pseudo}, we see that the energy loss vanishes both for $v\rightarrow v_\mt{max}$ and for $v\rightarrow v_\mt{lim}$. The former is due to the fact that the factor \eqn{sqrt} vanishes in this limit. The latter is implemented by the fact that $q_0^\mt{min}$ diverges as 
$v \rightarrow v_\mt{lim}$, which in turn is a manifestation of the Heaviside theta function in eq.~\eqn{scalarloss}. As reflected in figs.~\ref{result-scalar} and \ref{result-pseudo}, the maximum energy loss occurs at some intermediate velocity such that 
$\vlim < v_\mt{int} < v_\mt{max}$. Although $v_\mt{int}$ is not easy to compute, we know that $v_\mt{max}-v_\mt{lim}\propto z_0^2$. It then follows that also $v_\mt{max}-v_\mt{int} \propto z_0^2$ and therefore that the maximum energy loss diverges as $1/z_0^6$.

\subsection{Transverse vector mesons}
The energy lost into these modes is given by eq.~\eqn{transresult} which, utilizing \equ{hqfs}, leads to
\be
\frac{dE_\mt{trans}}{dt}= \frac{e^2 v}{2\Omega_3} \left(\beta^3\right)^2 \frac{\Omega^2}{2}
\int^\infty_{q_0^\mt{min}}  \frac{dq}{2\pi} q^3 \exp\left(-\frac{1}{2} \Omega q z^2_0\right) 
\left(1-\frac{v^2_\mt{lim}}{v^2}- \frac{2\Omega}{q v^2}\right) \,.
\ee
Unlike for scalar meson emission, it is easy to see that in this case the energy loss is a monotonically growing function of the velocity. Thus, the maximum value of the energy loss is attained for $v=v_\mt{max}$ and, to leading order in $z_0$, it is given by
\be
\label{elmaxT}
\frac{dE_\mt{trans}}{dt}=\frac{e^2 }{4\pi\Omega_3} \left(\beta^3\right)^2
\frac{76 \, \exp(-4)}{ v_\mt{lim} z^6_0}\,.
\ee
This shows the same divergence for small $r_0$ as in the case of scalar modes.

\subsection{Longitudinal vector mesons}
Inserting \equ{hqfs} into the expression for energy loss \eqn{dEdtVlong}, and to leading
order in $z_0$ we obtain
\be
\frac{dE_\mt{long}}{dt} = \frac{e^2 }{2\Omega_3}  \, \frac{ 2 \Omega^3}{ v} \frac{1}{\left(\beta^4\right)^2}
\int^\infty_{q_0^\mt{min}}   \frac{dq}{2\pi} q^2 \left(1-\frac{1}{8} \Omega q z_0^2\right)^2
\exp\left(-\frac{1}{2} \Omega q z^2_0\right) \,.
\ee
As in the transverse vector case, the maximum energy loss is attained at 
$v=v_\mt{max}$. Using \eqn{qogen} and setting $\ell^\alpha=1$
(see the definition below eq.~\eqn{eqho}) the maximum energy loss is
\be
\label{elmaxV}
\frac{dE_\mt{long}}{dt} = \frac{e^2}{4\pi \Omega_3} \,
\frac{1}{\left(\beta^4\right)^2} \frac{632 \exp(-6)}{v_\mt{lim}\, z_0^6}\, .
\ee
Again, this diverges with the same power of $z_0$ as in the cases of transverse and scalar modes. Note also that since $\beta^4=1/\beta^3$ and the numerical factors in eqs.~\equ{elmaxV} and \equ{elmaxT} are similar, we find that in this limit the energy lost into transverse and longitudinal vector modes is comparable.

Finally, we may comment on the mass dependence of the energy lost into vector mesons. The maximal energy loss is proportional to
\be
\left. \frac{d E}{d t} \right|_\mt{max}\propto \frac{R_0^4\left(R_0^4-1\right)^2}{\left(R_0^4+1\right)} \frac{1}{r^6_0}\,.
\ee
The fact that this is a growing function of $R_0$ means that the energy loss increases if the quark mass is increased while keeping all other parameters such as $r_0, v$, etc.  fixed. However, this should not be necessarily taken as an indication that Cherenkov energy loss in a real HIC experiment increases as the quark mass increases, because the `preferred' or `mean' values of these parameters with which a quark is produced may themselves depend on the quark mass.

\section{Low-temperature limit}
As $T \rightarrow 0$ with all other scales fixed, the redshift at the bottom of the branes decreases, and so the limiting velocity of mesons, $\vlim$, approaches unity. In turn, this means that the momentum $q_0 (T)$ at which Cherenkov radiation turns on diverges as $T \rightarrow 0$,
much in the same way as in the case $r_0 \rightarrow 0$ studied above.
The purpose of this section is to estimate $q_0(T)$ in the low-temperature limit.
As we will see, the product $q_0 (T) \,T^2$ remains finite as 
$T \rightarrow 0$, which makes this limit harder to study than the $r_0 \rightarrow 0$ limit.

As $T \rightarrow 0$ with fixed quark mass, the parameter $m \propto \mq/T$ controlling the asymptotic position of the branes, $R(r\rightarrow \infty) = m$, diverges. For this reason it is convenient to introduce rescaled coordinates 
\be
\hat x^\mu = \frac{x^\mu}{\sqrt{\eps}} \sac
\{ \hat r, \hat R, \hat \rho \} = \sqrt{\eps} \{r,R,\rho\} \,,
\label{rescale}
\ee
where $\eps \equiv 1 / m^2$ and we are interested in the limit $\eps \rightarrow 0$. The rescaling of the $x$-coordinates is chosen so that the induced metric on the D7-brane takes the same form as that in (\ref{metric}), i.e.~$ds^2=L^2 ds^2(g)$ with
\be
ds^2 (g) =  \frac{\hat \rho^2}{2} \left[-\frac{f^2}{\tilde f} d \hat t^2 + {\tilde f} d \vec {\hat x} ^2\right]
+ \frac{(1+\dot {\hat R} ^2)}{\hat \rho^2}  d \hat r^2+ \frac{\hat r^2}{\hat \rho^2} d \Omega^2_3 \,,
\ee
where now
\be
f(\hat \rho)= 1- \frac{\eps^2}{\hat \rho^4} \sac 
\tilde{f}(\hat \rho)=1+\frac{\eps^2}{\hat \rho^4} \,.
\label{ff}
\ee
In this new set of coordinates the horizon is located at $\hat \rho_\mt{hor} = \sqrt{\eps}$, whereas $\hat R (\hat r \rightarrow \infty) = 1$. In fact, in the limit  $\eps \rightarrow 0$ the gravitational pull of the black hole becomes very small and the brane bends very little. Inserting the ansatz $\hat R (\hat r) = 1 + \delta \hat R( \hat r)$ in eq.~\eqn{embedeq} and linearizing in $\delta \hat R$ one finds that $\delta \hat R= \mathcal{O}(\eps^4)$. Since we will work to order $\eps^2$ we will neglect this correction. Furthermore, it is easy to see that the equations of motion for the transverse (\ref{trans}), longitudinal (\ref{PHI}) and
scalar (\ref{X}) modes are unmodified by the rescaling (\ref{rescale}) provided
we rescale the momentum in a consistent way, i.e.~$\{\hat \omega, \hat q\} = \sqrt{\eps} \{\omega, q\}$, so that we leave the product $(\omega t - \vec q \cdot \vec x)$ invariant.
For ease of notation,  in the following we will omit the hat symbol.

Following appendix C, eqs.~(\ref{trans}), (\ref{PHI}) and (\ref{X}) can be rewritten in the Schr\"odinger form (\ref{potgeng}). The potential (\ref{potgen}) is in general a complicated function, but it simplifies in the limit $\eps \rightarrow 0$. Indeed, in this case eq.~(\ref{zorgen}) can be integrated explicitly with the result 
\be
z(r) = \sqrt{2} \arctan{r} + \frac{3}{8 \sqrt{2}} \left( 3 \arctan{r} + 
\frac{r (5 + 3 r^2)}{{(1 + r^2)}^2} \right) \eps^2 + \mathcal{O}(\eps^4) \,,
\label{zr}
\ee
so that $z_\mt{max} \equiv z(r \rightarrow \infty) \simeq \pi / \sqrt{2}$. The potential can then be written as
\be
V^\alpha(z,q) = V_1(z) q^2 + V{_2^\alpha}(z) \,,
\label{V}
\ee
where
\bea
V_1 (z) &=& 1 - 4 \cos^4 \left( \frac{z}{\sqrt{2}} \right) \eps^2 + g(z) \eps^4 + 
\mathcal{O} (\eps^6) \,,
\label{v1} \\
V{_2^\alpha} (z) &=& \frac{ \mathrm{A}^\alpha \cos^2 \left( \sqrt{2} z \right) + 
\mathrm{B}^\alpha \cos \left( \sqrt{2} z \right) +
\mathrm{C}^\alpha }{2 \sin^2 \left( \sqrt{2} z \right)} + \frac{(m^\alpha)^2}{1+\cos \left( \sqrt{2} z \right)} + h(z) \eps^2 +  \mathcal{O} (\eps^4) \,, \,\,\,\,\,\,\,\,\,\,\,
\label{v2}
\eea
with $\mathrm{A}^\alpha = \{ 0,0,1,1 \}$, $\mathrm{B}^\alpha = \{ -6,-6,0,8 \}$ and
$\mathrm{C}^\alpha = \{ 9,9,2,6 \}$. The functions $g(z)$ and $h(z)$ are smooth and bound, and their explicit form will not be needed. 

Before proceeding further let us clarify one point. In the limit $z \rightarrow z_\mt{max}$, $V{_2^4} (z)$ shows a negative divergence. This is related to the fact that the radial profile 
$\Psi^4 (r) = \Phi(r)$ does not vanish near the boundary. This may seem counterintuitive, since generically one expects a `confining' potential near the boundary (the AdS `box') which is partly responsible for the discreteness of the spectrum. However, note that the physical electric field $E$ is related to $\Phi$ through eq.~(\ref{Eophi}), which may be written as $E = a^4(z) \prt_z \Phi$. In the limit $r \rightarrow \infty$, the factor $f^2/\tilde{f}^2$ in eq.~(\ref{eqngen}) approaches 1, so one can differentiate this equation with $\alpha = 4$ and use the fact that $a^4(z) = 1/ a^3(z)$ to show that the asymptotic form of the equation for $E$ is identical to that for the transverse mode $\Psi^3 = \mathcal{A}$.

Let us now return to the potential \eqn{V}. As explained above, we are interested in the limit 
$\eps \rightarrow 0$ and $q \geq q_0 (\eps) \rightarrow \infty$. In particular, we wish to determine whether in this limit the product $\eps q_0$ goes to zero, remains finite,
or diverges. We will establish that the product remains finite by showing that
the other two possibilities lead to a contradiction. 

Consider first the possibility that $\eps q_0 \rightarrow \infty$ as 
$\eps \rightarrow 0$. Then the potential takes the form
\be
V^\alpha (z,q) - q^2 \simeq 
\frac{ \mathrm{A}^\alpha \cos^2 \left( \sqrt{2} z \right) + \mathrm{B}^\alpha \cos \left( \sqrt{2} z \right) +
\mathrm{C}^\alpha }{2 \sin^2 \left( \sqrt{2} z \right)} + \frac{(m^\alpha)^2}{1+\cos \left( \sqrt{2} z \right)} 
- 4 \cos^4 \left( \frac{z}{\sqrt{2}} \right) \eps^2 q^2 \,,
\label{potsimple}
\ee
where the last term dominates everywhere except near the endpoints $z=0$ and  
$z_\mt{max}$, at which the order-one part of the potential, given by the first two terms on the right-hand side, diverges. As an illustration, fig.~\ref{pot} shows the potential for the 
$\alpha=3$ - mode for several values of $\eps q$.~We see that the potential develops a minimum at a small value of $z$ as $\eps q$ becomes large. 
\FIGURE{
\,\,\,\,\,
\includegraphics[width=0.42 \textwidth]{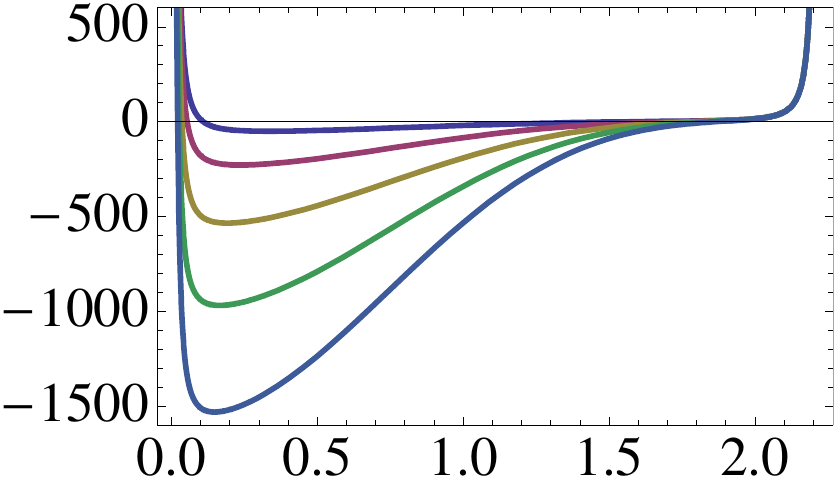} 
\put(-80,-10){$z$}
\put(-202,32){\rotatebox{90}{{$V^3(z,q)-q^2$}}}
\caption{Potential for $\alpha=3$, in the limit $\eps \rightarrow 0$, $q \rightarrow \infty$ 
and $\eps q$ fixed. From top to bottom the curves correspond to $\eps q = 4,8,12,16,20$.}
\vspace{2mm} 
\label{pot}} 
\noindent
This allows us to consider an expansion for small $z$ in order to obtain the energy levels associated to the potential. In this way we reproduce the result (\ref{drhq}), approximated for $\eps \rightarrow 0$. For example, in the case $\alpha=3$ and $n=0$ the dispersion relation takes the form:
\be
\omega(q) = \left( 1 - 2 \eps^2 \right) q + \frac{2}{\sqrt{5} q} \sqrt{1 + 20 q^2 \eps^2} + 
\mathcal{O} (\eps^2) \,.
\ee
The value of $q_0$ is determined by the condition $\omega (q_0) = v q_0$, where
$v$ is the quark velocity, which depends on the quark position through
\be
v(r_0) = 1 - \frac{2 \eps^2}{\rho^4_0} + \mathcal{O} (\eps^4) \,.
\ee
The result is 
\be
\eps q_0 = \left[ -10 + \sqrt{5} \frac{\sqrt{20 + r{_0^2} (2+r{_0^2}) \left( 40 + 21r{_0^2}(2+r{_0^2}) \right) } }{(1+r{_0^2})^2} \right]^{-1/2} \,.
\label{finite}
\ee
This expression is not parametrically large for any value of $r_0$ except in the limit $r_0 \rightarrow 0$, which was considered above and is unrelated to the zero-temperature limit under consideration here. In particular, for small $r_0$ eq.~\eqn{finite} yields $\eps q_0 = 1 / {r_0}^2 + 3/2 + \mathcal{O} (r_0^2)$, whereas for large $r_0$ one finds  
$\eps q_0 = 2.01 + 1.99/r_0^4 + \mathcal{O} (1/r_0^{6})$. We therefore conclude that 
$\eps q_0(\eps)$ remains finite as $\eps \rightarrow 0$, in contradiction with our initial assumption that $\eps q_0(\eps) \rightarrow \infty$ in this limit.

Consider now the opposite possibility, i.e.~that $\eps q_0 \rightarrow 0$ as 
$\eps \rightarrow 0$. In this case the last term in the potential \eqn{potsimple} is small and the energy levels can be determined using perturbation theory with the result
\be
\omega{_n^2} (q) = q^2 + \lambda_n - w_n \eps^2 q^2 + \mathcal{O} (\eps^4 q^4) \,.
\ee
Here, $\lambda_n$ are the eigenvalues of the problem in the absence of the perturbation, and $-\omega_n$ is the (negative) first-order correction given by the expectation value of the perturbation in the $n$-th eigenstate, $\langle n | -4\cos^4 (z / \sqrt{2})  |n \rangle$ . The key point is that $\lambda_n$ and $w_n$ are independent of $\eps$ and $q$. In order to find the crossing point, we need to solve $\omega{_n^2} (q_0) = v^2 q{_0^2}$. Since the minimum possible value of $q_0$ corresponds to the maximum quark velocity, $v=1$, we find
\be
\eps q_0 \simeq \sqrt{\frac{\lambda_n}{w_n}} \,,
\ee
which is in contradiction with our initial assumption that $\eps q_0 \rightarrow 0$. 

We therefore conclude that the combination
\be
\eps \hat q = \eps \sqrt{\eps} q = \frac{1}{m^3} \frac{\tilde q}{\pi T} = \left( \frac{\sqrt{\lambda}T}{2 \mq} \right)^3 \frac{\tilde q}{\pi T} = 
\left(\frac{2\pi T}{\mmes} \right)^3 \frac{\tilde q}{\pi T} 
\ee
remains finite in the limit $T \rightarrow 0$, where we have reinstated the hat and we recall that $\tilde{q}$ is the physical, dimensionful momentum. We see that in this limit
$\tilde q (T) \, T^2 \sim \mq^3 / \lambda^{3/2} \sim \mmes^3$, which remains finite in the low-temperature limit, as we anticipated. This means that the relevant potential in this limit is \eqn{potsimple}, where all terms are of the same order. This makes the problem harder than that associated to the limit $r_0 \rightarrow 0$, and we have been unable to find analytic expressions for the corresponding eigenfunctions and eigenvalues.

\end{document}